\documentclass[12pt,twoside]{article}

\usepackage{hyperref}
\usepackage{setspace}
\usepackage{indentfirst}
\usepackage{geometry}
\usepackage{xcolor}
\usepackage{fancyhdr}
\usepackage{caption}
\usepackage[labelfont=bf]{caption}
\usepackage{natbib}
\usepackage{caption}
\usepackage{subcaption}
\usepackage{graphicx}
\usepackage{floatrow}
\usepackage{booktabs}
\usepackage{tabularx}
\usepackage{titling}
\usepackage[hang,flushmargin]{footmisc}
\usepackage{titlesec}
\titleformat{\section}{\normalfont\bfseries\filcenter}{}{0pt}{}
\titleformat{\subsection}{\normalfont\bfseries\filcenter}{}{0pt}{\itshape}
\titleformat{\subsubsection}{\normalfont\bfseries\filcenter}{}{0pt}{\itshape}
\titlespacing*{\section}
  {0pt}{-.1\baselineskip}{-.1\baselineskip}  
\titlespacing*{\subsection}
   {0pt}{-.1\baselineskip}{-.1\baselineskip}  
\usepackage[T1]{fontenc}
\date{}
\providecommand{\keywords}[1]
{
   \small	
  \textit{\hspace{-1em} Keywords: } #1
}

\usepackage{bm}
\usepackage{amsmath,amsfonts,amsthm,amssymb}

\fancypagestyle{firstpage}{
  \fancyhf{}
}
\title{\normalsize \textbf{
Analyzing and Estimating Support for U.S. Presidential Candidates in Twitter Polls
} \vspace{-1.5em}} 

\author{
\normalsize STEPHEN SCARANO \\ 
\normalsize University of Massachusetts Amherst, U.S.A. 
\vspace{1em} 
\\ 
\normalsize VIJAYALAKSHMI VASUDEVAN\\ 
\normalsize University of Massachusetts Amherst, U.S.A. 
\\ 
\normalsize CHHANDAK BAGCHI\\ 
\normalsize University of Massachusetts Amherst, U.S.A. \\
\normalsize MATTIA SAMORY \\ 
\normalsize Sapienza University of Rome, Italy\\
\normalsize JUNGHWAN YANG \\ 
\normalsize University of Illinois Urbana-Champaign,  U.S.A.\\
\normalsize PRZEMYSLAW A. GRABOWICZ \\ 
\normalsize University of Massachusetts Amherst, U.S.A. 
\thanks{\hspace{-1em} Corresponding author: pgrabowicz@umass.edu} 
} 
\usepackage{lipsum}
\renewcommand\footnotemark{}
\begin{document}
\maketitle
\thispagestyle{firstpage}
\vspace{-8em}

\begin{abstract}
\noindent 
Polls posted on social media have emerged in recent years as an important tool for estimating public opinion, e.g., to gauge public support for business decisions and political candidates in national elections. Here, we examine nearly two thousand Twitter polls gauging support for U.S. presidential candidates during the 2016 and 2020 election campaigns. 
First, we describe the rapidly emerging prevalence of social polls.
Second, we characterize social polls in terms of their heterogeneity and response options.
Third, leveraging machine learning models for user attribute inference, we describe the demographics, political leanings, and other characteristics of the users who author and interact with social polls. 
Finally, we study the relationship between social poll results, their attributes, and the characteristics of users interacting with them.
Our findings reveal that Twitter polls are biased in various ways, starting from the position of the presidential candidates among the poll options to biases in demographic attributes and poll results. 
The 2016 and 2020 polls were predominantly crafted by older males and manifested a pronounced bias favoring candidate Donald Trump, in contrast to traditional surveys, which favored Democratic candidates. We further identify and explore the potential reasons for such biases in social polling and discuss their potential repercussions. Finally, we show that biases in social media polls can be corrected via regression and poststratification. The errors of the resulting election estimates can be as low as $1\%$-$2\%$, suggesting that social media polls can become a promising source of information about public opinion.

\end{abstract}

\keywords{\textit{public opinion, opinion polls, social media} \vspace{8ex}}

\section{Introduction}
Social media and the Internet provide an unprecedented opportunity to observe political dialogue on a large scale. Platforms such as Twitter and Facebook have transformed users from passive consumers into active participants who produce and share information. This shift has led to a phenomenon where the buzz generated by users on social media not only serves as an indicator of public interest in various topics but also can potentially influence mainstream media coverage \citep{Wells2016}. Consequently, social media has garnered significant attention from the public, media, and political elites, who are seeking to gauge what people think about key issues in society~\citep{mcgregor2019social}.

Over the last decade, social media have emerged as an important source of political information~\citep{masonwalker2021news, Wells2016, mcgregor2019social}. Journalists have employed social media in various ways, from capturing public responses to events such as debates and assessing presidential candidate performances~\citep{mcgregor2019social} to amplifying the reach of political messages from Donald J. Trump by featuring his tweets in mainstream news~\citep{Wells2016}. For the general public, social media is not only a platform for community building but also serves as a tool for political engagement and voter mobilization~\citep{bond201261millionperson, rojas2009mobilizers}. It is also an educational tool on political and social issues~\citep{perrin202023, bail2021breaking}.
Moreover, the influence of social signals, such as user responses and comments, on shaping public perceptions cannot be understated~\citep{muchnik2013social, 
kramer2014experimental}.
These pivotal roles social media plays in current politics underscore the need to examine the expressions of public opinion and discourse on social media.

Researchers have been exploring various methods to systematically estimate public opinion using social media data\citep{Beauchamp2017, DiGrazia2013, Zhang2022, Li2023}. For one example, Zhang et al. \citeyearpar{Zhang2022} tracked textual contents expressed by sub-groups on Twitter to understand how politically opposing groups react differently to particular political events. Other studies used similar techniques to estimate the results of traditional opinion polls from each state \citep{Beauchamp2017} and to approximate politicians' electoral performances \citep{DiGrazia2013}. However, these methods have encountered limitations in accuracy and predictive capability, arguably due to the complexity of inferring unambiguous signals of political support from the natural language content of tweets, or indirect signals like re-sharing~\citep{oconnor2010tweets, jungherr2015, schober2016, klasnja2018measuring, dong2021review}.

Our study focuses on understanding a more direct indicator of public opinion on social media: \textit{social polls}. These polls, widespread on social platforms, are informal surveys created and shared by users. Elon Musk, the owner of X (formerly known as Twitter) famously used this feature to make key business decisions, e.g., about Twitter's CEO change \citep{Mehta2022}. Due to their ease of use and quick turnaround time for gauging users' support for political candidates, Twitter polls have gained popularity during election campaigns and it is not uncommon to see social polls amass hundreds of thousands of votes. Such polls can be regarded as a more direct method of sensing public opinion compared to earlier techniques that inferred it from the text of tweets or user engagement through retweets and mentions. Yet, little scholarly attention has been paid to understanding social polls. 


Social polls encode political information that users express, endorse, and share spontaneously, and thus convey a novel form of political engagement that traditional surveys often miss. 
The unique affordances of social media, such as visibility, editability, persistence, and association \citep{Treem2013}, that are inherent to networked technologies, can facilitate the amplification and spread of both information and social actions. This new way of social engagement can potentially challenge traditional norms of social interaction and affect the discourses in online public spheres\citep{boyd2011}. In particular, for individuals critical of the elite-dominated political culture and mainstream media, the affordances of social media offer opportunities to create and spread their narratives. Social media platforms emerge as critical arenas where users can express their viewpoints and, importantly, construct and reinforce their social identities \citep{boyd2011}.

However, we acknowledge the limitations of social polls as proxies for public opinion. Social media data often disproportionately reflect the expressed views of a reactive, polarized, and highly engaged subset of users \citep{Zhang2022} and can be influenced by the activity of bots and astroturfing accounts \citep{keller2020campaign, Ferrara2016}. Despite their resemblance to traditional surveys, social polls inherently lack scientific rigor due to non-systematic sampling and the absence of demographic information of the respondents, leading to potential biases in poll outcomes \citep{Auxier2021, wojcik2019twitterpop}. Thus, while social polls provide a rich insight into political behaviors on social media, they should be taken with caution when considered as indicators of broader public opinion. 

In this study, we systematically describe social polls, their integrity, and the extent of their biases, to clarify their relevance in the online political landscape and their idiosyncrasies compared to traditional surveys. 
Our analyses focus on a specific set of Twitter polls posted during the concluding months of the 2016 and 2020 U.S. presidential elections (e.g., ``Who has your vote? Biden or Trump?'' and ``Would you vote for Clinton or Trump in the upcoming election?''). 
We begin by describing Twitter polls related to the U.S. presidential elections. Then, we address the following research questions.

\noindent\textit{RQ1: What is the prevalence of social polls and how does it vary over time?}\\
\noindent\textit{RQ2: What are the characteristics of social polls?}\\
\noindent\textit{RQ3: What are the characteristics of users who engaged in social polling?}\\
\noindent\textit{RQ4: What relationships exist between these characteristics and poll outcomes?}\\
\noindent\textit{RQ5: Can biases in social poll outcomes be reduced to mine public opinion?}

To answer these research questions, we assess the extent of bias in Twitter polls by comparing their results with those from traditional election polls and exit polls. Our examination focuses on potential sources for the discrepancy between Twitter polls and actual election results, considering factors such as (1) order effects of vote options in Twitter polls, (2) the lack of partisan and (3) demographic representativeness among potential respondents of the polls, and (4) the presence of bots interacting with these polls. Finally, we show that the discrepancy found in Twitter polls can be reduced by correcting for the sources of bias in poll outcomes, which pioneers an alternative path toward the prospect of mining public opinion from social media data. 

\section{Related Work}
\label{sec:rel-work}


\subsection{Social polls as political engagement}
\label{related-work-socpol}

Public opinion, defined as an aggregate of individual opinions \citep{Price1992}, is a crucial part of a well-functioning democracy. During the election, understanding public opinion becomes even more important for campaigners as it offers insight into the voters' support towards political candidates and informs campaign strategies. 
One of the most popular ways to assess public opinion is survey-based opinion polls \citep{Price1992}. 
However, skepticism about the efficacy of polling methods dates back to 1948, when sociologist \citet{blumer1948} raised concerns about the limitations of survey methods in capturing the true essence of public sentiments. 
Critics point to the influence of interactions among diverse groups within society in forming public opinion~\citep{blumer1948, fishkin2006}, arguing that surveys fail to consider the societal hierarchy and the influence of key individuals and groups in opinion formation~\citep{blumer1948, herbst1998}. Other scholars point out the fundamental limitations of opinion polls by arguing that survey-based public opinion deviates from fundamental democratic ideals, such as active participation, open discussion, and thoughtful deliberation~\citep{berelson1952, habermas1989}.



On the other hand, social polls can be regarded as indicators of political engagement within specific subsets of the population, rather than as accurate reflections of the general public opinion. 
This approach suggests that the observed opinions on social media are inherently biased, irrespective of the volume of trending messages because those who actively post political content on these platforms are more likely to have strong political views. Thus, it will be challenging if we try to reproduce the results of traditional opinion polls with social media data. A historical example illustrating the pitfalls of relying on a biased sample is the Literary Digest's 1936 survey. This survey aimed to forecast the outcome of the U.S. Presidential election using 10 million ballots but failed despite receiving 2.3 million responses. The failure was mainly due to a low response rate and a bias in the demographic of respondents who returned their ballots \citep{Squire1988}. In a similar vein, the vast number of tweets generated on Twitter every hour should not be misconstrued as a comprehensive representation of the wider public opinion, as they are likely biased towards the views of a more politically vocal subset of Twitter users. Thus, in this study, we conceptualize social polls as signals of political engagement rather than an accurate reflection of (traditional) public opinion.
Our inaugural research question asks about the prevalence of this emergent form of political engagement.

\noindent\textit{RQ1: What is the prevalence of social polls and how does it vary over time?}

The new way of political engagement on social media can be attributed to their unique technological affordances. Social media affordances are the perceived and potential properties of social media that emerge from the interplay of technology, social dynamics, and contextual factors \citep{Ronzhyn2022}. As these properties both enable and define specific uses of these platforms, communication scholars categorize social media as a type of networked publics \citep{boyd2011}. This perspective highlights that the distinctive technological affordances of these platforms enable a variety of actions, such as one-click retweeting of user-generated content and nurturing the formation of connections and networks among users \citep{Treem2013} and they can shape social media as a venue where they create and share political information and political actions that potentially lead to online public spheres  \citep{Ronzhyn2022, boyd2011}. The unique affordances of social media, especially Twitter polls, may shape user responses and their answers to the polls. To further examine the relationship between various affordances of Twitter polls and the votes, we dissect our remaining research questions into nuanced inquiries.

\noindent\textit{RQ2.1: Are there correlations between the number of votes and the counts of retweets and favorites?}

Based on our discussion above, we argue that social media users may perceive platforms like Twitter, particularly Twitter polls, as a means of expressing their political views and influencing political discourse. In particular, for those critical of the elite-dominated political culture and mainstream media or for those who have conspiratorial beliefs, these affordances of social media present opportunities to construct and propagate alternative narratives and perspectives. This pattern has been documented in a recent study done by the Pew Research Center. The study reveals systematic differences in the trust of Republicans and Democrats in different types of media. Democrats trust news media more if they perceive them as ``mainstream'', whereas Republicans if \textit{not} mainstream~\citep{gottfried2021republicans}. Similarly, there is a trend that individuals identifying as young Republicans trust social media more than national news media, while young Democrats trust more national news media than social media~\citep{liedke2022us}.
One may expect that these differences in trust are reflected in the ways that Republican and Democrat users create and interact with political polls on Twitter. Thus, we propose RQ 2.2 to examine whether the positioning of response options of Twitter polls reflects the biases of poll authors.

\noindent\textit{RQ2.2: How are the voting options presented in social polls?}

\subsubsection{Biases in Twitter poll participants} \label{related-work-dem}


In general, social media users in the U.S. tend to skew more liberal in their political affiliations compared to the general population. A Pew Research Center study shows that 36\% identify as Democrats among social media users, whereas the corresponding figure in the U.S. general population is 30\% \citep{wojcik2019twitterpop}. Moreover, the observable behavior of these populations may differ. Research suggests that conservatives on social media are more inclined to engage with liberal discourse on contentious issues such as gun control, compared to their liberal counterparts \citep{Zhang2022shooting}. Our next research question aims to delve into the political leanings of individuals participating in Twitter polls, seeking to uncover any potential political biases within them.

\noindent\textit{RQ3.1: What are the political orientations of users who participate in social polling?}

On a similar note, social media users do not accurately reflect the demographic makeup and location distribution of the general population. Audiences on Twitter and Facebook are significantly younger~\citep{mellon2017population, wojcik2019twitterpop}. Facebook users skew towards a female demographic, while Twitter users are biased towards men \citep{mislove2011understanding, mellon2017population, wojcik2019twitterpop}. To examine these biases in Twitter polls, we propose RQ 3.2.

\noindent\textit{RQ3.2: What are the demographic characteristics of users who participate in social polling on Twitter?}




Finally, social media data could be influenced by inauthentic activities such as astroturfing campaigns and bots. As famously manifested in Russia's \textit{Internet Research Agency} (IRA) intervention in the 2016 U.S. Presidential election, many organized or sponsored inauthentic social media operations are uncovered and documented \citep{Schoch2022}. Similarly, research has long found that a vast amount of bot accounts generate social media posts daily, and, further, that the prevalence of bot accounts on Twitter has only increased since Elon Musk's acquisition of the site in 2022 \citep{Ferrara2016, hickey2023auditing}. 
The tools for manipulating social media discourse, now broadly accessible via online services, have expanded their reach beyond governmental institutions to everyday consumers \citep{alrawi2020ira}. These external influences have the potential to skew metrics and distort the representation of public opinion, particularly as some campaigns with political motivations deliberately exploit them to sway the course of social media dialogues \citep{Schoch2022}.
If these kinds of inauthentic information operations and bot activities can influence Twitter activities, they can also create another bias in Twitter polls. Thus, we aim to identify the extent of inauthentic accounts related to Twitter Polls.

\noindent\textit{RQ3.3: What is the fraction of bot accounts among these users?}

\subsection{Biases in Twitter poll outcomes}

The biases in polls and characteristics of users engaging with Twitter polls could result in biased poll outcomes.
To begin with, the positioning of vote options in such polls may be biased and this bias can influence social poll results, similarly to how it influences traditional poll results~\citep{Strack1992}.
Further, the opinions of specific demographics can be over-represented because 
(i) Twitter users are more likely to be male and young \citep{mislove2011understanding, wojcik2019twitterpop}, 
and (ii) politically-interested users are non-representative of all users \citep{hughes2019national,hughes2021small}.
Finally, the prevalence of bot accounts, astroturfing campaigns, and artificial likes and comments, might easily distort various metrics from Twitter including social polls \citep{keller2020campaign, Ferrara2016}. With the following research questions, our study aims to provide detailed descriptions of the poll outcomes and their biases.

\noindent\textit{RQ4.1: How do the results of social polls compare to those of mainstream polls?}\\
\noindent\textit{RQ4.2: How do poll attributes, such as the order or response options, relate to their outcomes?}\\
\noindent\textit{RQ4.3: How do user attributes relate to the outcomes of social polls?}



\subsection{Mining public opinion from social media} \label{related-work-socmed}

%
Researchers have attempted to use unconventional data to interpolate public opinion. One of the popular ideas is to use social media data to capture public opinion \citep{oconnor2010tweets, DiGrazia2013, jungherr2015, schober2016, Beauchamp2017, dong2021review}. The large-scale data from social media not only reveals communication patterns and political preferences of individuals but also can present a tremendous scientific opportunity to understand individual and group behavior in real time across various issues. If sufficiently accurate, this method could complement mainstream polls in gauging public opinion. 
Prior research demonstrated limited viability of non-representative social media data in accurately estimating state-level public opinion \citep{Beauchamp2017} and in getting unique insights into how different sub-groups publicly express their political opinions \citep{Zhang2022}. 
The limitations of such approaches stem from the challenge of extracting and aggregating opinions from natural language, which we largely sidestep by focusing on social polls, and from the challenge of accounting for various biases present in social media data~\citep{klasnja2018measuring}.
However, \citet{Wang2015} demonstrated that with the appropriate statistical adjustments of demographic and political variables, non-representative polls from Xbox users can yield accurate election forecasts. These findings underline the potential of unconventional data sources in capturing public opinion once biases in data are corrected. Following this promise, we address our final research question.

\noindent\textit{RQ5: Can biases in social poll outcomes be reduced to mine public opinion?}

By addressing this question, the study contributes to the ongoing discussion about the role and reliability of social media as a tool for gauging public opinion.

\section{Data}

\begin{figure}[t!]
	\centering
	\includegraphics[width=0.99\columnwidth]{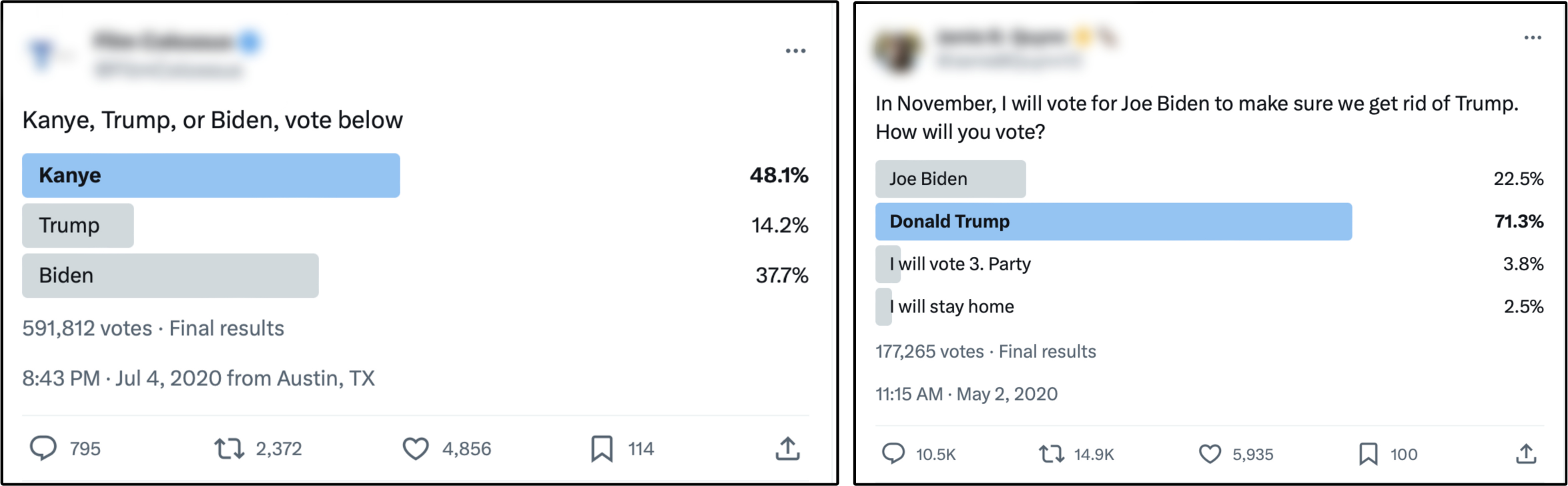}
	\caption{The two Twitter polls with the highest number of votes in the dataset.}
	\label{fig:poll-examples}
\end{figure}

We collected and labeled a large dataset of Twitter polls gauging support for the 2016 and 2020 presidential elections and users interacting with these polls. To facilitate the reproduction of the results of this study, we share the tweet IDs of all collected polls and user IDs of all users interacting with these polls on Zenodo\footnote{Link will be provided once this study is publicly released.}.

\paragraph{Social polls}
Provided access to the Twitter API v2, we compiled election polls from the periods of 11/02/2019--11/30/2020 and 11/02/2015--11/30/2016 by making full-archive searches for the queries ``\texttt{vote trump biden}'' and ``\texttt{vote trump clinton}'' respectively. 
This step resulted in a vast number of polls. 
To identify polls gauging support for presidential candidates (Donald Trump and Hillary Clinton in 2016, and Donald Trump and Joe Biden in 2020), we focused on the polls that explicitly mention them in the post's text. This resulted in 4,551 polls.
To identify relevant polls, we manually inspected all of them. To this end, we developed a labeling guideline that defines as relevant the polls gauging support for the 2016 and 2020 presidential election candidates either by (i) directly asking for voting preferences (e.g., "Who has your vote?"), or (ii) asking for election predictions (e.g., "Who do you think will win the presidential election vote?").
This approach identified a total of 1,753 relevant Twitter polls when excluding polls posted after respective election days (see examples in Figure~\ref{fig:poll-examples}).
A subset of $194$ polls was labeled by two trained coders. They have achieved on this set of items an almost perfect inter-rater agreement, measured with Cohen's kappa, of $0.914$ $(p<0.001)$.



\paragraph{User profiles and reactions to social polls}
Using Twitter API, we pulled each poll's author profile as well as those of the corresponding followers. Then, we collected all retweeters and favoriters of each poll. Finally, we pulled followees of poll authors, retweeters, and favoriters, to conduct political affiliation inference.

\paragraph{Mainstream polls}
In addition to the Twitter set, we collected 192 mainstream polls from 2020 and 144 from 2016 aggregated by \textit{FiveThirtyEight}~\footnote{\url{https://projects.fivethirtyeight.com/polls}} and used national exit poll data distributed by the \textit{Roper Center} \citep{NEP2016, NEP2020}.

\section{Methods}

We augmented the profiles of poll authors and their followers, as well as poll retweeters and favoriters, by inferring their attributes (age, gender, political affiliation, botness, organization status) using machine learning models and by extracting location information directly from their profiles. To verify the accuracy of the inference, we asked two trained annotators to label a subset of polls and users.

To compare the results of polls with more response options than the two main presidential candidates, we normalized poll outcomes by dropping the extra options and normalizing the fraction of votes for the two main presidential candidates. Then, we relate potential sources of bias to poll outcomes using a regression model. Finally, we poststratify the outcomes of social media polls using the inferred user attributes and regression models.

\paragraph{Gender and Age}
We classified the gender and age of users, attributes commonly understood to correlate with voting behavior, and used to stratify or poststratify mainstream surveys~\citep{silver2021death}. To do so, we employ the multilingual, multimodal, and multi-label machine learning tool \textit{M3-Inference} \citep{wang2019inference}. \textit{M3-Inference} is a deep learning text and image model that uses usernames, profiles, and photos to infer age and gender with state-of-the-art accuracy while diminishing algorithmic bias in comparison to other approaches.
Since the model additionally infers the likelihood that the given account represents an organization, we exclude from our analysis those users who exceed an org score of $0.90$.

\paragraph{Location}
Participation in Twitter polls may differ by location, both between U.S. States and globally. To this end, we inferred the location of Twitter users. On Twitter, users can optionally disclose their position, in plain text, using the \texttt{location} field of their profile. We resolved such entries to geolocations using Photon, an open-source geocoder built for OpenStreetMap data.\footnote{\url{https://photon.komoot.io/}} We could infer the location of 4,737,715 users thusly. To expand coverage, for users whose location could not be geocoded via the previous method, we combined the \texttt{location} and \texttt{description} plain-text fields of user profiles, and extracted emojis corresponding to national flags. After excluding cases of users displaying flags of multiple countries, we could infer the location of an additional 156,788 users via this second method. In total, 4,894,503 users were mapped to countries, 1,019,610 of which could be resolved to specific U.S. States.

\paragraph{Political Ideology}
We estimate relative political ideology in Twitter polls using a Markov chain Monte-Carlo approach by the authors of \citep{barbera2015tweeting, Barbera2015}. 
The tool infers the political ideology of a user based on a set of users who that user follows (so-called \textit{followees}); that is, if user $A$ in the majority follows well-known right-wing accounts and user $B$ follows well-known left-wing accounts, the tool will output a positive value for user $A$ and a negative value for user $B$. 
More specifically, a user following \texttt{@RealDonaldTrump} and \texttt{@FoxNews} may be classified as rightward (> 0), while an equivalent user following \texttt{@JoeBiden} and \texttt{@CNN} may be classified as leftward (< 0). 
Each user instance is mapped to a continuous political ideology
value in the interval $[-3, 3]$.
While we provide the raw distribution of these values in Section 3, we later discretize this range into three bins (\textit{Left}, \textit{Moderate}, \textit{Right}), splitting the space evenly for simplicity. Prior work by \citep{barbera2015tweeting} shows that this approach performs comparably with standard ideological assessment surveys. 
%

The model infers users' political affiliation from their followees---the users an account is following. However, due to the rate limitations of the Twitter API, collecting the followees for the millions of accounts in this study would have been prohibitive. Thus, we infer political affiliation for all poll authors, retweeters, and favoriters, and for a random subset of followers. 

\paragraph{Botness}

We subjected authors of both the 2016 and 2020 election polls through the machine learning classifier \textit{Botometer} to estimate the distribution of human and bot accounts. Trained on a dataset of 5.6 million tweets, \textit{Botometer} is a random forest classifier that evaluates network, user, friend, temporal, content, and sentiment features to label a profile as authentic or artificial \citep{davis2016botornot}. 
\textit{Botometer} provides a bot score within the interval $[0, 1]$ to gauge bot likelihood, but it does not specify a cut-off threshold for determining whether an account is a bot or not. After annotating a subset of users, we can determine a candidate threshold by ensuring that a human annotator and Botometer classify a similar fraction of users as bots. In this way, we set the cut-off threshold to $0.83$. 
We obtain bot scores for all poll authors and retweeters and for $35\%$ of favoriters and 6\% of author followers. We were not able to get bot scores for the remaining users, since the Botometer service ceased its operation once Twitter stopped providing free API access to researchers in the middle of 2023 when this study was conducted.

\paragraph{Validation of polls and inferred attributes}

Although the algorithms we used for estimating demographic and political attributes have undergone prior testing and validation, given the heavy reliance of our study on social poll data from Twitter and on estimates of various demographic and political attributes, we took additional steps to validate the relevancy of the data we used and the results obtained from machine learning algorithms. To accomplish this, we rely on the judgments of human coders to manually validate the integrity of the poll data and the inferred attributes.

The first validation task determines whether the data--a collection of Twitter polls--are relevant to the theoretical concept that we aim to measure--public opinion on election outcomes. To accomplish this task, three coders independently rated whether the collected poll data were relevant to our research focus. The main criteria for this process are as follows: the poll should at least include both of the top leading candidates of each election (i.e., Hillary Clinton and Donald Trump in 2016 and Joe Biden and Donald Trump in 2020) and the poll instructions directly ask about users' voting intention or the winning projection of the elections. Out of more than 4,000 polls that we collected from Twitter API, we selected 1,950 polls that were deemed to be relevant to the study. Then, we constructed a random sample of 196 polls and asked the coders to code them independently. The three coders' inter-rater reliability (IRR) measured in Fleiss' Kappa was 0.914. Based on this initial coding, around 39\% of the polls in the sample are classified as relevant. Given the high IRR, one of the coders coded the rest of the dataset and identified a final 1,950 relevant polls.

The second validation task involves evaluating the precision of estimating the demographic and political attributes of Twitter users, which were used in this study to describe potential biases in Twitter polls. Specifically, we focus on validating the following variables: (1) distinguishing between organizational and personal accounts, (2) discerning bot-like or human-like traits of the account, (3) estimation of the political leaning of the account, and for personal accounts, identifying the (4) age and (5) gender of the account holder. Since we are comparing the outcomes of machine learning and human judgment on the outcome, we asked one human coder to review the estimated outcome of those four attributes from a random set of 239 Twitter accounts. To simplify the process, we showed the machine-determined attributes to the coder (not a co-author) and asked them to determine whether they agreed with those classifications or not. The results are promising. Based on the coder's validation, our methods achieved approximately 93\% accuracy in distinguishing between organizational and personal accounts, 91\% accuracy in assessing bot-likeness, 93\% accuracy in estimating political ideology, 91\% accuracy in estimating the age of the account holders, and 88\% accuracy in classifying the gender of the account holders.

\begin{figure}
    \centering
    \includegraphics[width=0.8\columnwidth] {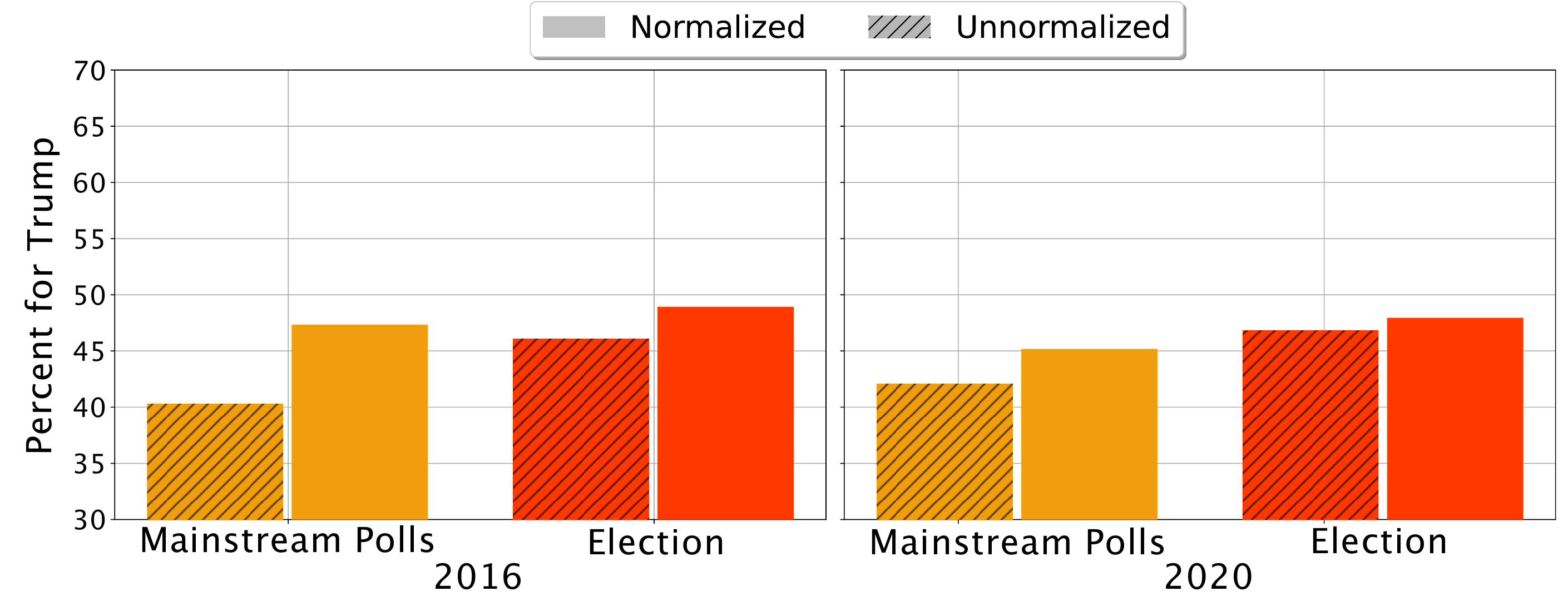}
    \caption{A comparison of the raw (hatched bars) and normalized (bars without hatching) outcomes of mainstream polls and respective presidential elections.
    We unify all polls and election outcomes to focus on the head-to-head race between the two main presidential election candidates by dropping votes for other candidates and response options and normalizing the fractions of responses for the two main candidates. This simple operation reduces the gap between the average mainstream poll outcomes (orange) and election outcomes (red), i.e., the orange and red bars without hatching have similar heights.
    }
    \label{fig:norm-bar}
\end{figure}

\paragraph{Normalized poll and election outcomes}
Many election polls, including social media polls, traditional polls, and election outcomes, offer respondents options other than the two main presidential election candidates. 
When comparing the results of polls offering two response options (for the main presidential candidates) with the results of polls with more options, we neglect the votes for the options other than the two main candidates (Trump vs. Clinton, Trump vs. Biden) and normalize the fraction of the votes for the main candidates so that they sum up to $100\%$ (hatched bars in Figure~\ref{fig:norm-bar}). Surprisingly, this simple procedure helps the accuracy of traditional polls by shifting them closer to the (normalized) election outcomes (compare same-color bars in Figure~\ref{fig:norm-bar}).

\paragraph{Regression of poll outcomes} 

To relate potential sources of bias to poll outcomes, we perform ordinary least squares regressions of poll outcomes, $y$, operationalized as the fraction of votes cast in favor of Trump in the 2016 and 2020 election seasons. As the potential sources of bias in poll outcomes we consider biases in the characteristics of potential poll respondents, including gender, age, political ideology, location, and authenticity. As proxies for voters, who are unknown since polls on Twitter are anonymous, we use retweeters and favoriters. 
Then, as predictors, we use marginal fractions $p^d_p(g)$ of potential voters in poll $p$ belonging to population stratum $g$. The resulting model is
\begin{equation}
\hat{y}_p = \sum_{d\in\mathcal{D}} \sum_{g\in \mathcal{G}_d \setminus \{g_d^\text{rf}\} } \beta^d_{g} p^d_p(g) + \beta_0,
\label{eq:linreg-polls}
\end{equation}
where 
$\mathcal{G}_d$ stands for all population strata within the dimension $d\in\mathcal{D}$, e.g., $\mathcal{G}_\text{gender}=\{\text{male},\text{female}\}$. 
The dimensions, $\mathcal{D}$, are used as predictors in the regression model and include several sociodemographic characteristics: the users' gender (male or \textit{female}), age group (\textit{less than 30}, between 30 and 39, greater than 40), political ideology (democrat, \textit{moderate}, republican), and location (U.S. red state, U.S. blue state, \textit{U.S. swing state}).
We also use bot classification outcome (bot or \textit{not bot}) as another feature to check whether the presence of bots is related to poll outcomes. Our final predictor encodes whether Trump or the Democratic candidate is listed first among poll response options (Trump or \textit{not Trump}).
The above sum excludes the reference strata, $g_d^\text{rf}$, marked in \textit{italics}, to avoid feature colinearity arising from the probability normalization $\sum_{g\in \mathcal{G}_d} p^d_p(g) = 1$.
To reduce noise in the dependent variable and the number of missing predictors, we exclude polls with fewer than $M=50$ votes.
We impute missing values in the remaining polls substituting them with the mean of the corresponding predictor.

\paragraph{Model-based poststratification} 

Multilevel regression and poststratification, pioneered by \citet{gelman1997postratification}, has propelled the study of public opinion and representation in subnational politics, evident in the proliferation of studies on public opinion and representation in leading academic journals~\citep{caughey2019public, caughey2022dynamic}.
Regression and poststratification comprise two steps. 
First, a regression is performed, like the one we introduced above. 
To increase the reliability of these models, we include only the characteristics that yield significant coefficients in at least one of the election years, i.e., $\mathcal{D}=\{$gender, age, political ideology, location$\}$.\footnote{Based on Results for RQ4.3.} 
Then, we poststratify social poll outcomes using these models.
In the second step, opinion estimates for a particular population are derived via the poststratification formula~\citep{jagers1986poststratification, little1993poststratification, smith1991poststratification, gelman2000poststratification}, which in our case is a weighted average of the learned coefficients,
\begin{equation}
    \hat{y}_e =\sum_{d\in\mathcal{D}} \sum_{g\in \mathcal{G}_d \setminus \{g_d^\text{rf}\} } \beta^d_{g} p^d_e(g) + \beta_0,
    \label{eq:poststrat}
\end{equation} 
where the weights, $p^d_e(g)$, are representative distributions of voters of category $g$ in election~$e$. Here, as $p^d_e(g)$ we use the 2016 and 2020 exit poll distributions of voter characteristics~$\mathcal{D}$, since exit polls employ a more robust data collection methodology and are more representative of the voting population during elections than Twitter polls. 

To make comparisons of election outcome estimates at the stratum level, we compute an estimate of the outcome of the election for the stratum $g$, 
\begin{equation}
    \hat{y}_e(g) =\sum_{d\in\mathcal{D}} \sum_{g'\in \mathcal{G}_d \setminus \{g_d^\text{rf}\} } \beta^d_{g'} p^d_e(g'|g) + \beta_0,
    \label{eq:poststrat}
\end{equation} 
where $p^d_e(g'|g)$ is the distribution over the strata $g'\in \mathcal{G}_d$. For the dimension $d$ that $g$ belongs to, $p^d_e(g'|g)$ becomes an indicator function $\mathbf{1}_{g}(g')$, i.e., if we condition on political ideology to be Democrat, then the distribution $p^d_e(g'|g)$ of political ideology is 1 for $g'=\text{Democrat}$ and 0 for other political ideologies. The value of the indicator function $\mathbf{1}_{g}(g')$ for $g'=g$ is 1, whereas for other $g'\neq g$ such that $g'\in \mathcal{G}_d$ it is 0.

\section{Results}

Next, we address each of our four major research questions.

\subsection{The prevalence of social polls}
\label{sec:rq1}



\begin{figure}[t!]
	\centering
	\includegraphics[width=0.99\columnwidth]{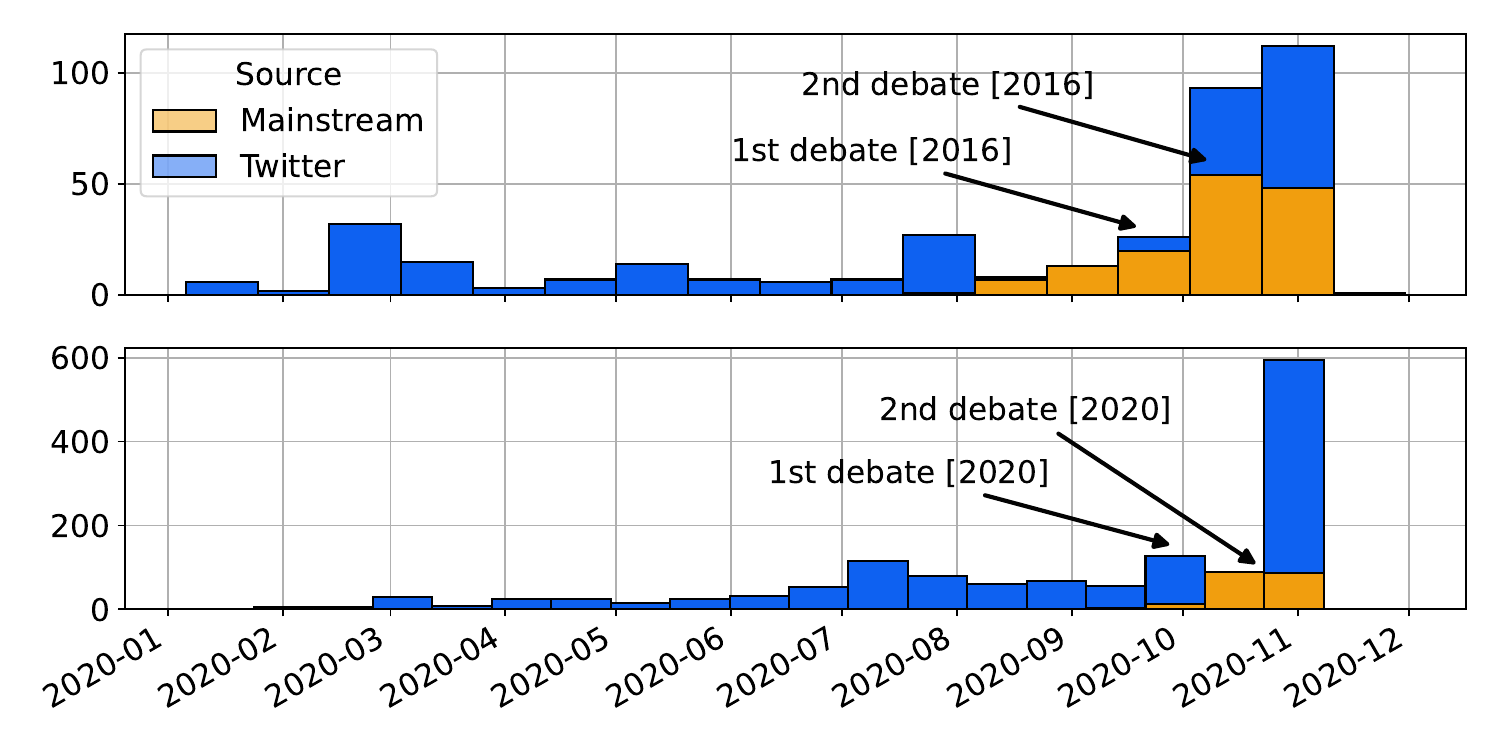}
	\caption{The number of Twitter and mainstream polls published throughout the 2016 (top) and 2020 (bottom) U.S. presidential election years.}
	\label{fig:polls_time}
\end{figure}

We first examined the prevalence of social polls and how it varies over time during the two election campaign periods (RQ1). To get a better understanding of this pattern, we visualize the number of social polls and mainstream polls side by side. Figure~\ref{fig:polls_time} suggests that both social and mainstream polls exhibit similar temporal trends, with an increase in the number of polls as election day approaches. The number of both polls peak in the last week of October for both 2016 and 2020. Notably, social polls have a broader temporal range, spanning almost the entire election year. Before September of the respective election years, there were 680 (32\%) social polls, while there were only 15 (4\%) corresponding mainstream polls in 2016 and none in 2020. Although the sample of mainstream polls from \textit{FiveThirtyEight} might not encompass all pertinent mainstream polls, there still is a strong indication that social polls are significantly more frequent between January and September 2020 than mainstream polls.

\begin{figure}[t!]
	\centering
	\includegraphics[width=0.8\columnwidth]{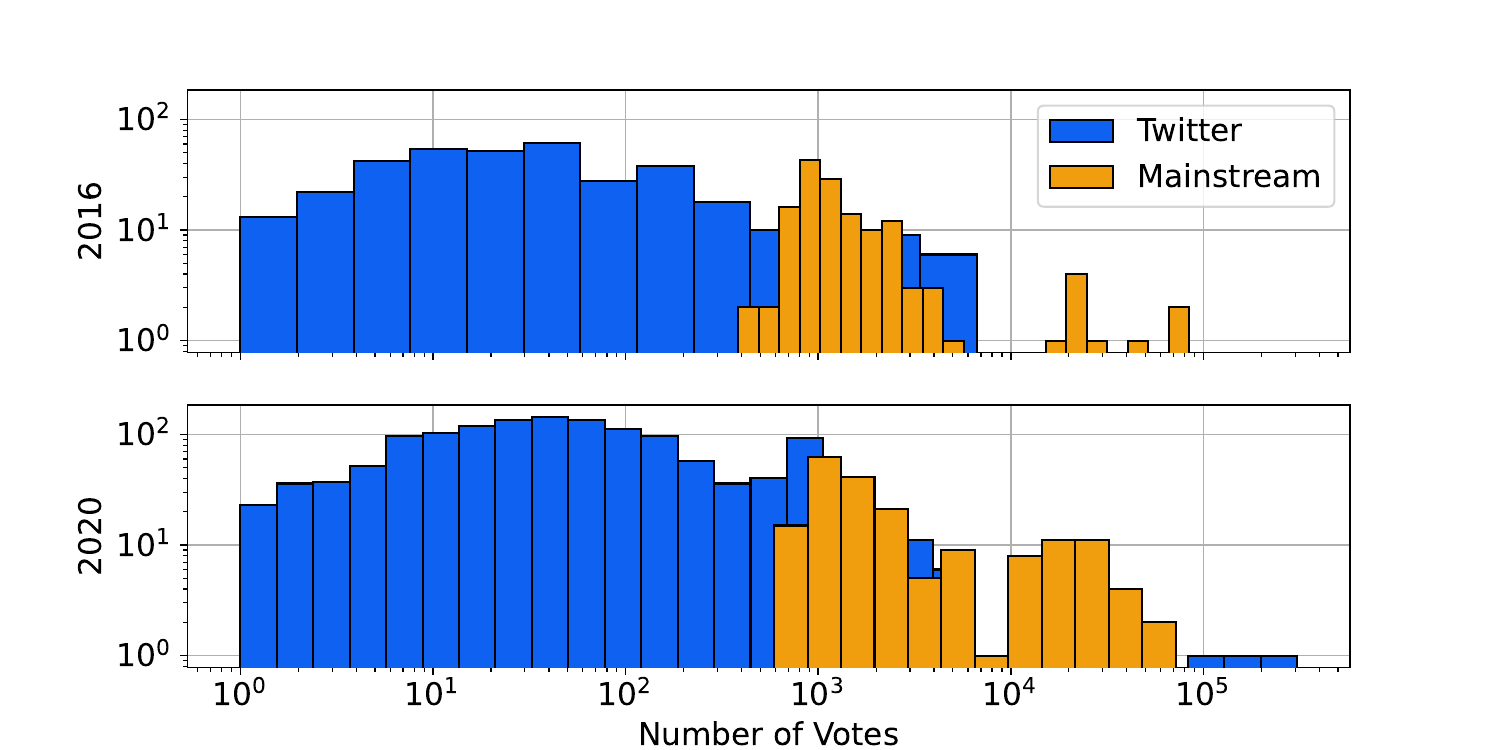}
	\caption{Distributions of the number of votes in Twitter and mainstream polls gauging support for the 2016 (top) and 2020 (bottom) U.S. presidential candidates.}
	\label{fig:vote_hist}
\end{figure}

In addition, we examine the number of votes that these social polls have gotten. As described in Table~\ref{tab:quant-summary}, social polls not only garnered a considerable number of votes, but also elicited significant user engagement through actions such as favorites, retweets, and followings. When visualizing the average number of votes (see Figure~\ref{fig:vote_hist}), it becomes apparent that the popularity of social polls assessing support for U.S. presidential candidates follows a heavy-tailed distribution. Although the majority (70\%) of polls receive fewer than $100$ votes, the top 15\% polls, in terms of vote count, often attracted comparable or sometimes many times larger numbers of votes than mainstream political polls.

It is important to note that our sample of polls is a small subset of all polls mentioning presidential election candidates among response options. In 2020 over 100.000 such polls were amassing 20 million votes, according to an estimate based on different datasets.\footnote{We will add the citation at the time of publication.} While many of these polls asked different questions than the ones in our dataset, social polls may be a treasure trove of public opinions.

Moreover, there was a substantial increase in social poll participation from 2016 to 2020. The total number of votes for social polls showed an eight-fold increase between the two election cycles as detailed in Table~\ref{tab:quant-summary}. 
%
If this trend persists, it is anticipated that the 2024 presidential election will witness a proliferation of social polls.
\begin{table}[t!]
    \centering
    \begin{tabular}{| c || c | c | c | c | c |}
    \hline
              & \multicolumn{2}{c}{\textbf{2016}}  &  \multicolumn{2}{|c|}{\textbf{2020}} & \textbf{Total} \\ 
              \hline
              & Total & Top 15\% & Total & Top 15\% & \\
        \hline
        \textbf{Votes} & 88,300 & 75,919 & 925,421 & 853,633 & 1,013,721 \\
        \textbf{Retweeters} & 2,991 & 2,669 & 36,290 & 34,216 & 39,281 \\
        \textbf{Favorites} & 2,623 & 1,915 & 26,078 & 23,519 & 29,514\\
        \textbf{Followers} & 450,612 & 140,712 & 17,728,247 & 12,853,901 & 18,178,859 \\
        \hline
        \textbf{Polls} & \multicolumn{2}{c}{\textbf{348}} & \multicolumn{2}{|c|}{\textbf{1,405}} & \textbf{1,753} \\
        \textbf{Authors} & \multicolumn{2}{c}{\textbf{298}} & \multicolumn{2}{|c|}{\textbf{960}} & \textbf{1,258}\\
        \hline
         
    \end{tabular}
    \caption{Quantitative summary of the dataset of Twitter polls gauging support for the 2016 and 2020 U.S. presidential candidates, including the number of polls and votes in these polls, as well as the number of poll authors, their followers, and the retweeters and favoriters of polls. All presented user counts are the numbers of unique users.}
    \label{tab:quant-summary}
\end{table}

\subsection{The characteristics of social polls}
\label{sec:rq2}

\begin{figure}[t!]
	\centering

    \begin{subfigure}[b]{0.35\textwidth}
         \centering
         \includegraphics[width=\textwidth]{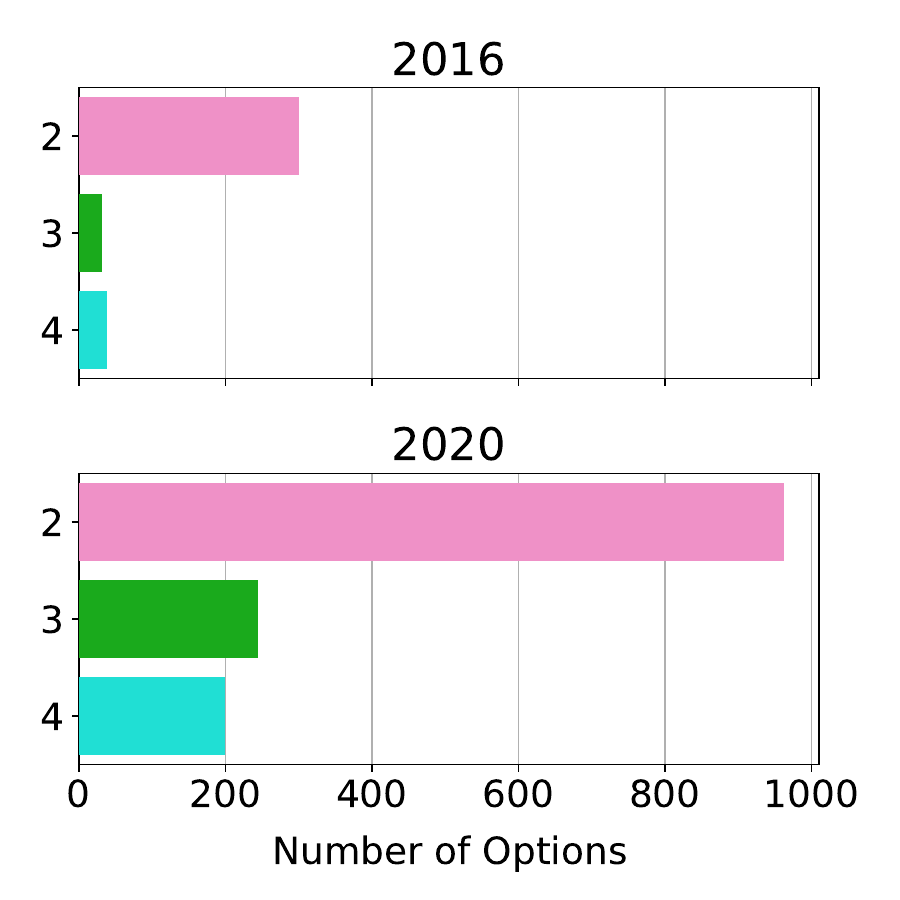}
         \caption{Histogram of the number of poll options.}
         \label{fig:option_bar}
     \end{subfigure}
    \hspace{1cm}
     \begin{subfigure}[b]{0.35\textwidth}
         \centering
         \includegraphics[width=\textwidth]{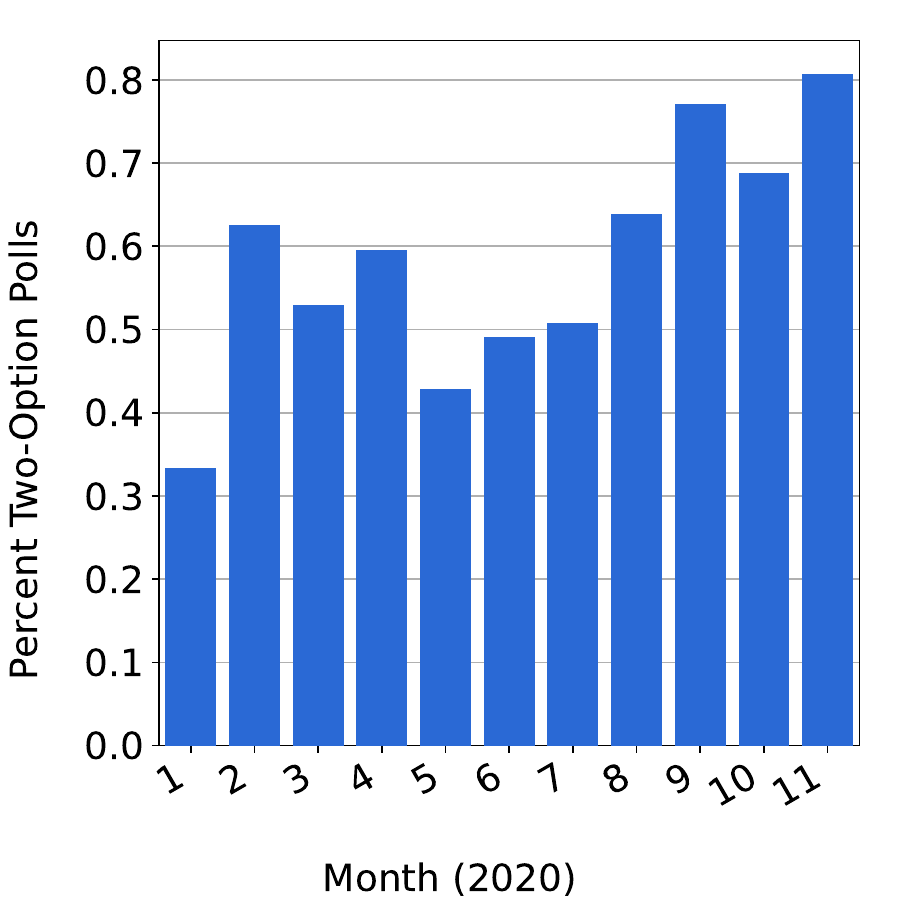}
         \caption{The fraction of head-to-head polls grows over the course of the election year.}
         \label{fig:option_temp}
     \end{subfigure}

     \caption{Breakdown of the number of Twitter poll options.}
	
\end{figure}

We turn our attention to describing various characteristics of social polls, focusing on user engagement and poll design (RQ2).

\noindent\textbf{RQ2.1}: Are there correlations between the number of votes and the counts of retweets and favorites?

\begin{figure}[t!]
	\centering
	\includegraphics[width=0.7\columnwidth]{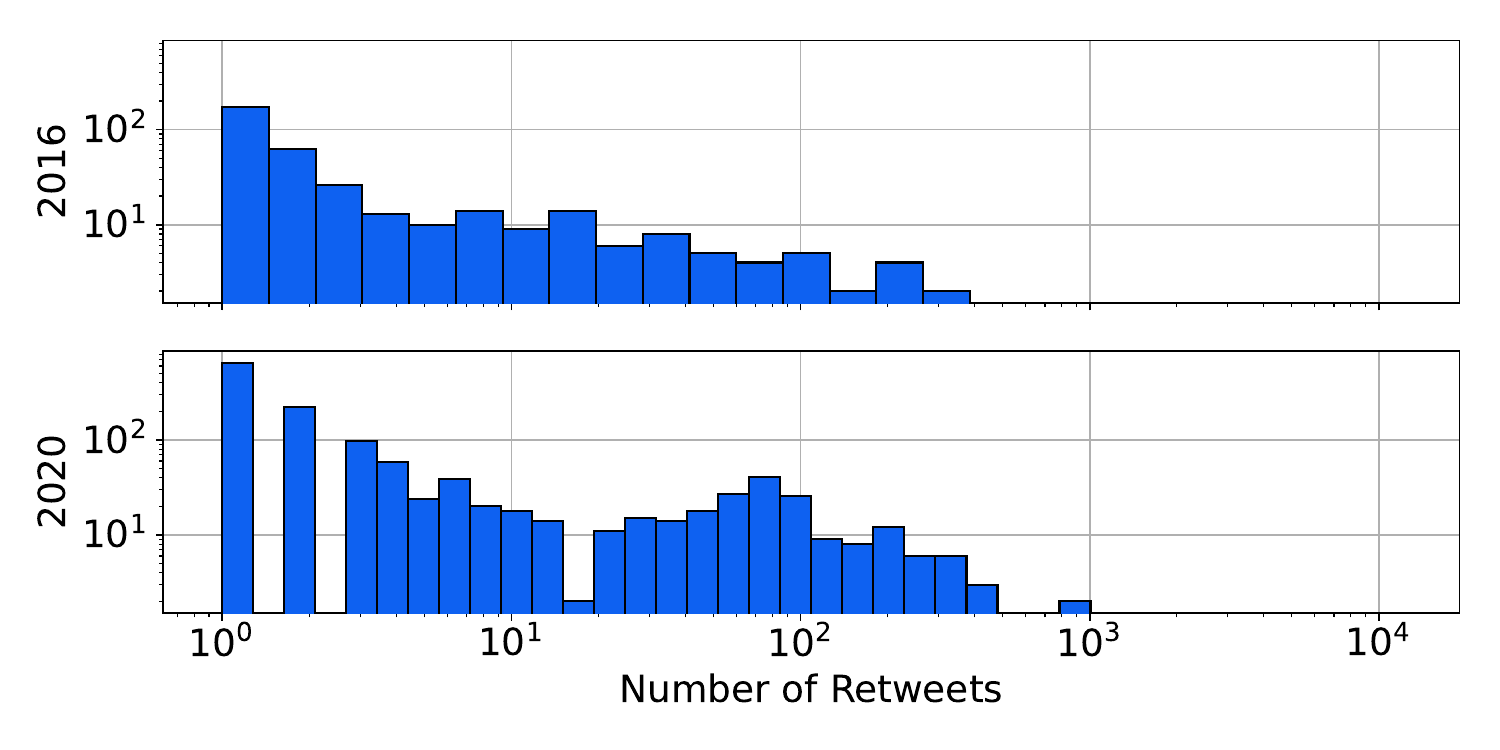}
	\caption{Histograms of the number of retweets per poll.}
	\label{fig:retweet-hist}
\end{figure}

\begin{figure}[t!]
	\centering
	\includegraphics[width=0.7\columnwidth]{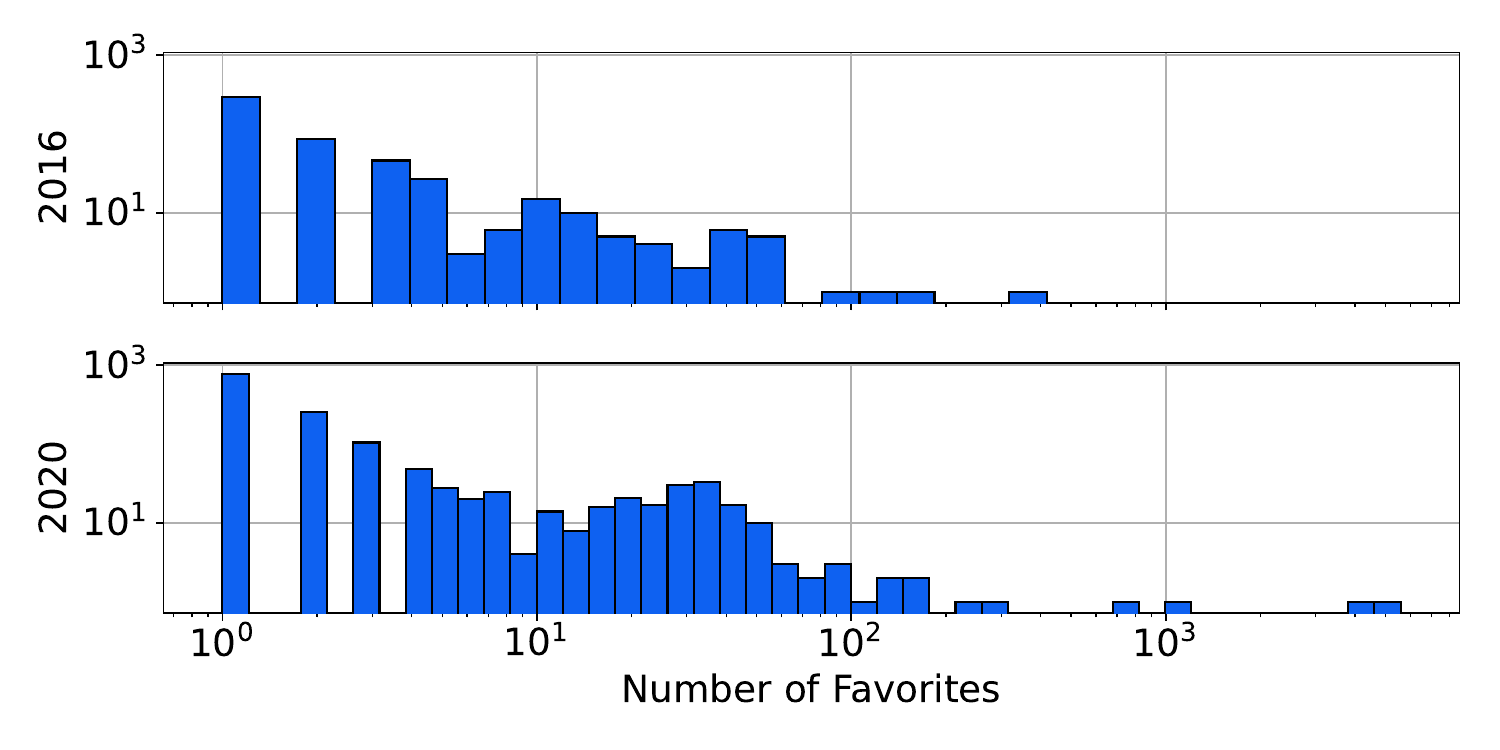}
	\caption{Histograms of the number of favorites per poll.}
	\label{fig:favorite-hist}
\end{figure}

One of the unique aspects of social polls is that people can engage with the polls by sharing the poll with their followers or by favoriting them to increase visibility. Figures \ref{fig:retweet-hist} and \ref{fig:favorite-hist} show that user engagements follow a heavy-tail distribution: 36,885 of the total 39,281 collected retweets (93\%) belong to the top 15\% of polls. Likewise, 25,434 of the total of 28,701 favorites (89\%) are concentrated within this same fraction (see Table~\ref{tab:quant-summary}). This suggests that only a relatively small fraction of social polls garner significant engagement from social media users. 

How do these user engagement metrics correlate with the number of votes? Retweets spread polls to new users, hence exposing them to more potential voters. Similarly, favorites reflect user interests. A high favorite count may influence other potential voters, possibly alike, to vote due to social influence~\citep{muchnik2013social, kramer2014experimental, grabowicz2020bayesian} or by boosting their exposure via Twitter's algorithmic news feed~\citep{huszar2022algorithmic}. Thus, it is plausible to suggest that higher retweet and favorite scores for tweets containing a poll can boost its visibility, thereby increasing the likelihood of further interactions with the poll.

We found that there are strong correlations between the number of votes and the number of retweets. Pearson's $r$ is $0.84$ $(p<0.001)$ for 2016 and $0.60$ $(p<0.001)$ for 2020, suggesting that polls attract more votes via retweets. We find similar correlations between the number of votes and the number of favorites: $r = 0.84$ $(p<0.001)$ for 2016 and $r = 0.86$ $(p<0.001)$ for 2020. 

These results suggest that there are strong positive correlations between user engagement and the number of votes. However, we do not yet know whether retweets push biased political polls into echo chambers, and whether this process exacerbates biases in poll outcomes. We will examine these questions in RQ3 and RQ4, respectively.

\noindent\textbf{RQ2.2}: How are the voting options presented in social polls?

The selection of response options in a survey has important implications for respondents' behavior and the quality of data. Research has found that the choice of response is related to response bias, response variability, and social desirability bias \citep{Price1992}. To answer RQ2.1, we describe the voting options provided in social polls.

By design, all social polls in our dataset must include the two leading candidates of each election campaign (Trump \& Clinton in 2016 and Trump \& Biden in 2020); however, some polls also include other candidates, providing up to four options to choose from. As shown in Figure~\ref{fig:option_bar}, the vast majority (72\%) of the dataset are 2-option polls, whereas 3-option and 4-option polls are less frequent (15\% and 13\%). 
As the presidential election campaigns develop and the Republican and Democratic party candidates emerge from the presidential primaries over the course of an election year, the fraction of the head-to-head polls increases (see Figure~\ref{fig:option_temp} for 2020, data for 2016 was too sparse to observe this pattern). 

\begin{figure}[t!]
	\centering
	\includegraphics[width=0.99\columnwidth]{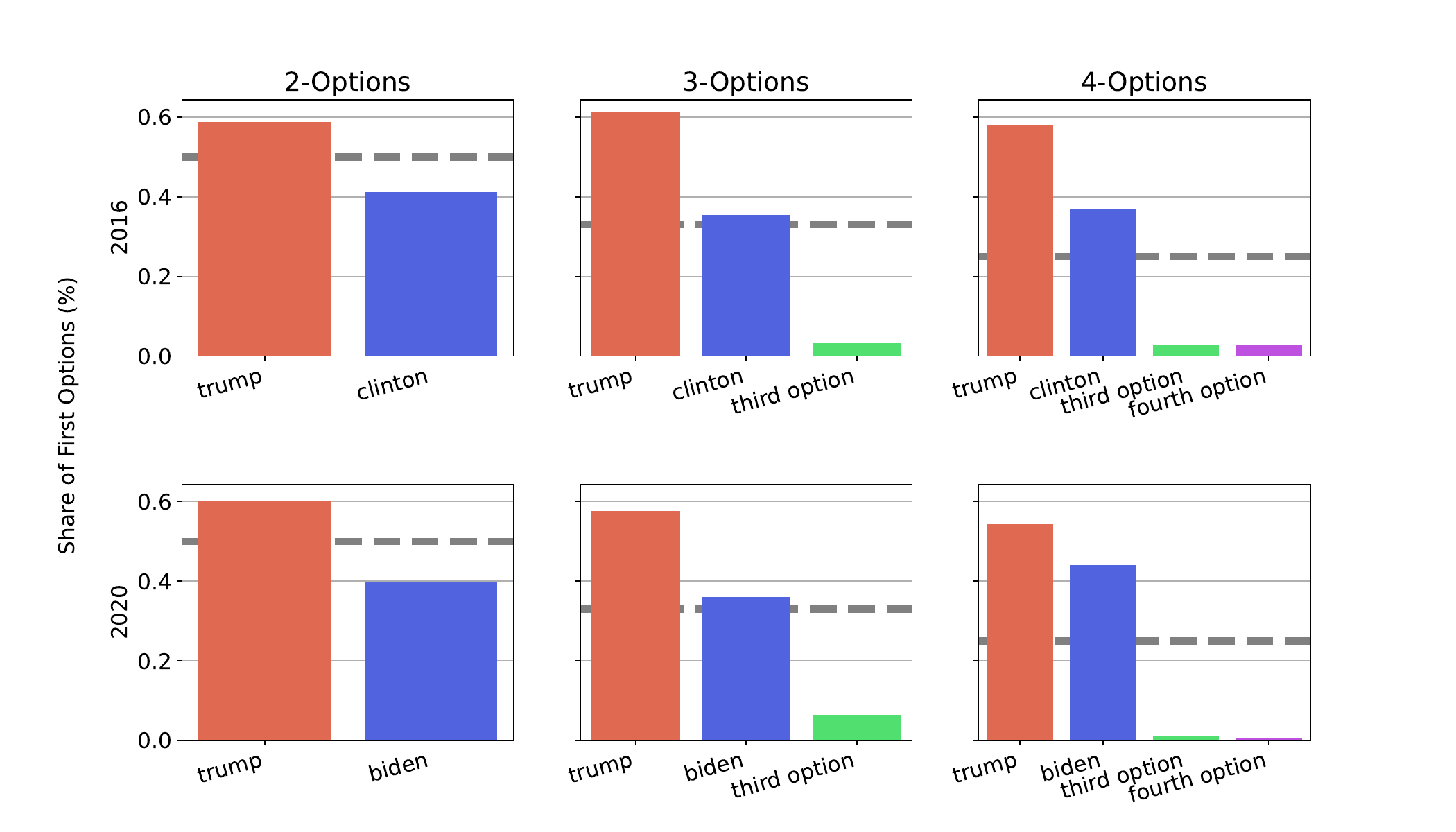}
	\caption{The fraction of Twitter polls listing a given candidate as first among potential answers. The majority of polls rank Trump as first. The dashed lines mark the fractions of $50\%$ for 2-option polls, $33.3\%$ for 3-option polls, and $25\%$ for 4-option polls, which correspond to the hypothetical condition where all candidates are ordered randomly (fairly) among poll options.}
	\label{fig:option_pos_bar}
\end{figure}

Another interesting aspect of polls to consider is the ``order effect,'' which is a well-documented phenomenon that different orders in response options can influence survey outcomes \citep{Strack1992}. Our data, shown in Figure~\ref{fig:option_pos_bar}, reveal the pattern where candidates from the two major parties consistently dominate the first and second option placements. Trump and the respective Democratic candidate consistently are consistently ranked first more often than the statistical expectations (i.e., 33\% for 3-option polls and 25\% for 4-option polls). Notably, regardless of the number of response options, Trump tends to be positioned above the Democratic candidate, suggesting a potential bias in poll design. We will further investigate this bias in response options' impact on poll outcomes in RQ4.1. 

\subsection{The characteristics of users engaged in social polling}


As discussed in Section \ref{sec:rel-work}, social poll outcomes may be the result of a non-representative user base engaged in the polling process. In this section, we outline the characteristics of the authors and potential voters in such polls. Considering that voters in social polls can encompass those who retweet or favorite the poll, as well as the followers of the poll authors, we discuss the attributes of these three groups of users. 
We assess whether the accounts are human versus automated bots and determine whether they resemble a representative sample in terms of their age, gender, and political orientation. These findings are then juxtaposed with data from corresponding exit polls.




\noindent\textbf{RQ3.1}: What are the political orientations of users who participate in social polling?

\begin{figure}[t!]
	\centering
	\includegraphics[width=0.99\columnwidth]{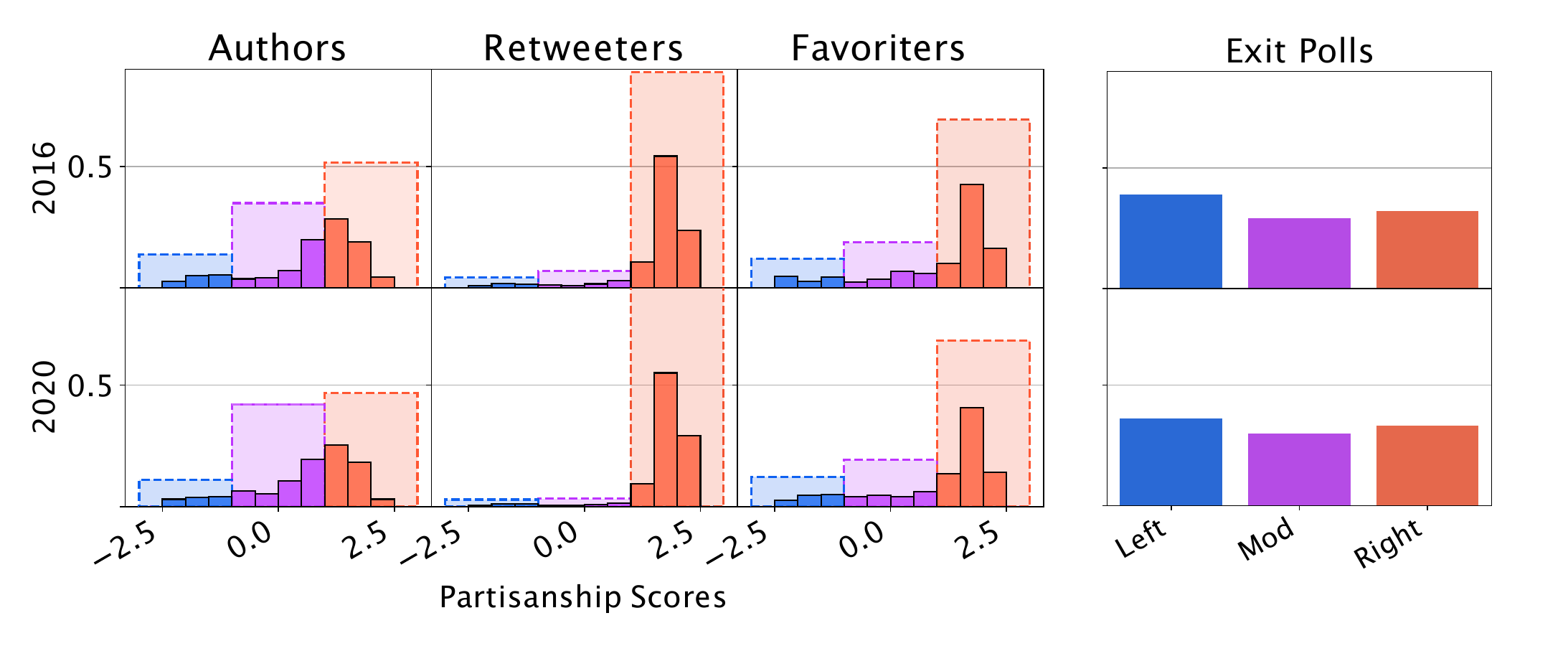}
	\caption{Political affiliation distribution of poll authors, retweeters, and favoriters. The applied political affiliation inference tool projects users into a continuous partisanship interval $[-3, 3]$, which we discretize into equally-sized partisanship bins (dashed bars of 3 distinct colors) to make comparisons to exit polls, which represent ideology using 3 categories: left, moderate, right.
 }
	\label{fig:demo-political}
\end{figure}

One of the most pertinent voter characteristics to political polls is the ideological makeup of the participants. We estimate the political ideology of the poll authors, retweeters, and favoriters using methods suggested by \citet{barbera2015tweeting}. Although an in-depth discussion of this method is out of the scope of the current paper, a brief explanation is provided in the Methods section. 

To compare Twitter users' political ideology scores with the political affiliation data from exit polls, we converted the continuous scale of inferred political ideology of users into a discrete scale. Also, for the purposes of this discussion, we use the terms political ideology and political affiliation interchangeably. The result of this classification is summarized in Figure~\ref{fig:demo-political}.
Our analysis shows that the distributions of political ideology of poll authors and users interacting with polls are skewed towards the right. This result is consistent with the fact that social poll results are skewed towards Trump (RQ2.2). Interestingly, retweeters and favoriters of social polls are even more likely to be more conservative than the poll authors themselves. This asymmetry resembles the one observed in the distribution of political ideology of users who interact with misinformation, which also predominantly leans to the right~\citep{nikolov2021right, gonzalezbailon2023asymmetric}.

\noindent\textbf{RQ3.2}: What are the demographic characteristics of users who participate in social polling?

\begin{figure}[t!]
	\centering
	\includegraphics[width=0.89\columnwidth]{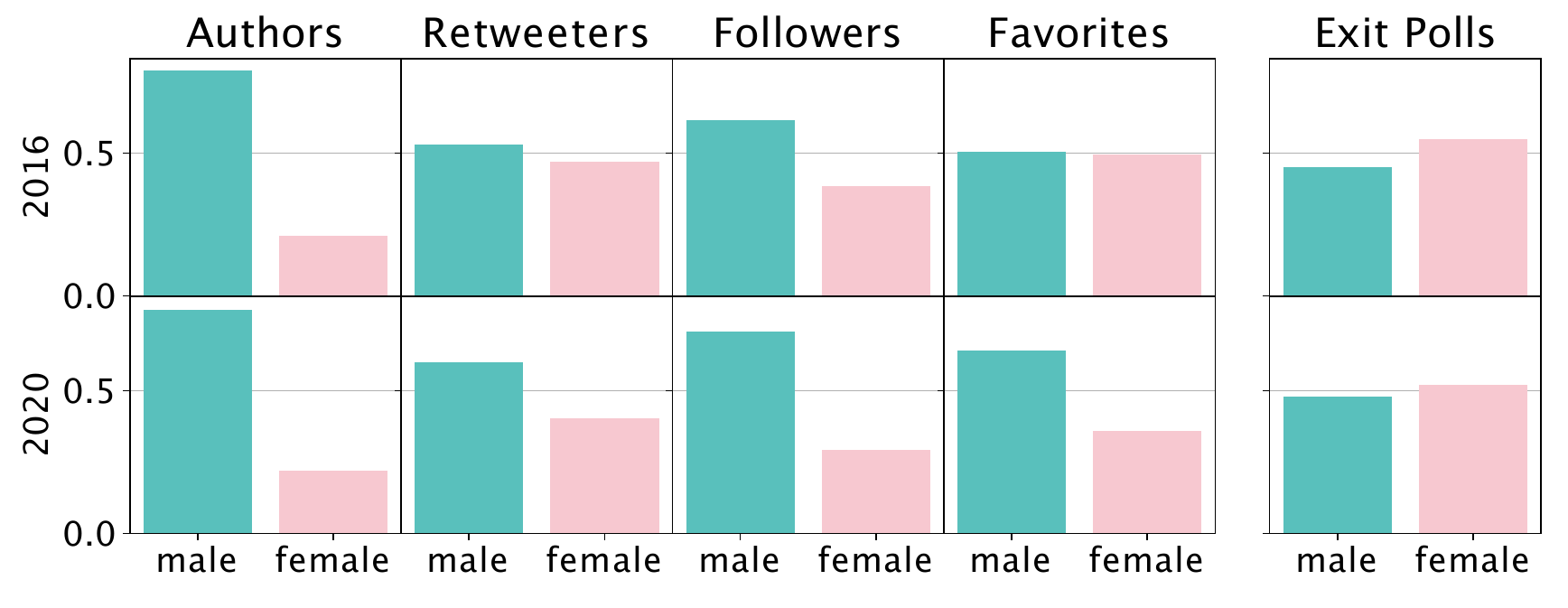}
	\caption{Gender distribution among social poll authors, their followers, as well as the retweeters and favoriters of the polls. For comparison, the rightmost figure shows gender distribution for the exit polls of 2016 and 2020, respectively.}
	\label{fig:demo-gender}
\end{figure}

\begin{figure}[t!]
	\centering
	\includegraphics[width=0.89\columnwidth]{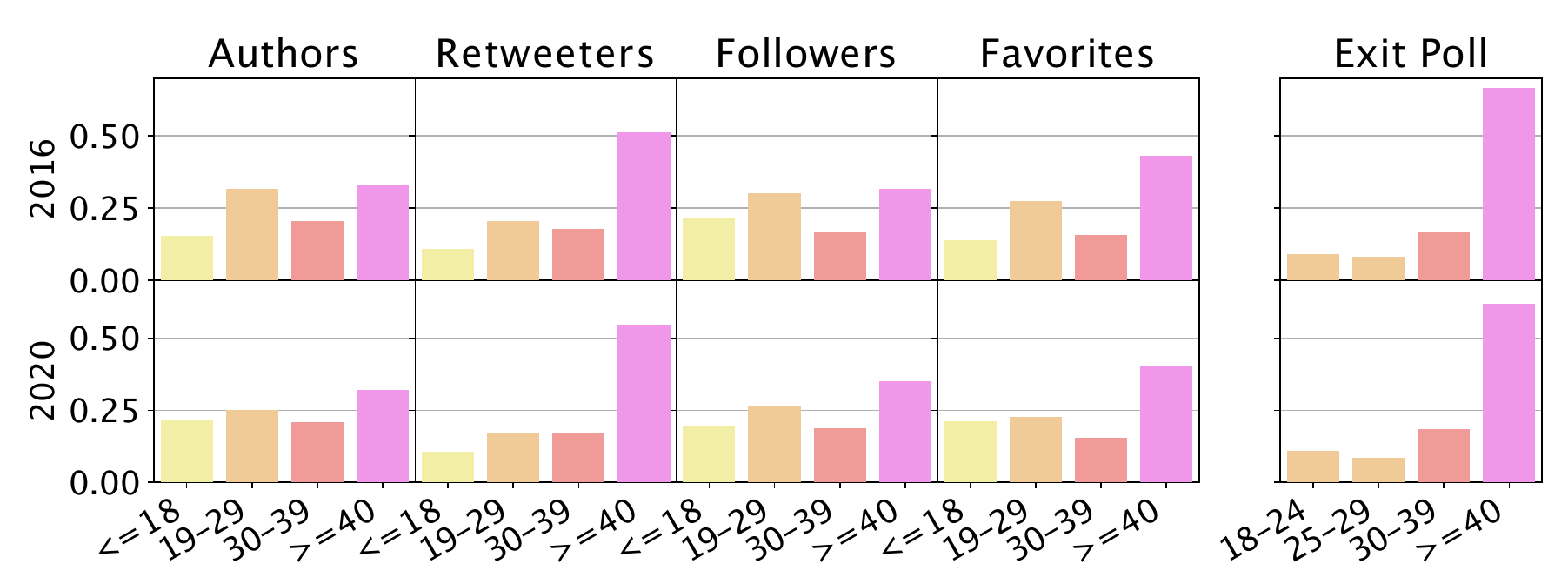}
	\caption{Age distribution of social poll authors, their followers, as well as the retweeters and favoriters of the polls. For comparison, the rightmost figure shows age distribution for the exit polls of 2016 and 2020, respectively. The bars are color-coded to mark the correspondence between the age brackets, e.g., the second bin for social polls corresponds to the first two age bins for exit polls.}
	\label{fig:demo-age}
\end{figure}

We analyze the gender and age of poll authors, retweeters and favoriters, and author followers. First, we find that the fraction of males is about 2 times larger among poll authors than among exit poll respondents (Figure~\ref{fig:demo-gender}), and that the fraction of poll authors below 30 years old is almost 3 times larger (Figure~\ref{fig:demo-age}). Our results conform to prior research suggesting that Twitter skews heavily male and young \citep{mellon2017population}. Interestingly, these biases are greater among authors and their followers than among retweeters and favoriters, suggesting that while young males are mobilized to author polls, people engaging with the polls are more similar to the general population in terms of age and gender.


\noindent\textbf{RQ3.3}: What is the fraction of bot accounts among users who participate in social polling?

\begin{figure}[t!]
	\centering
	\includegraphics[width=0.6\columnwidth]{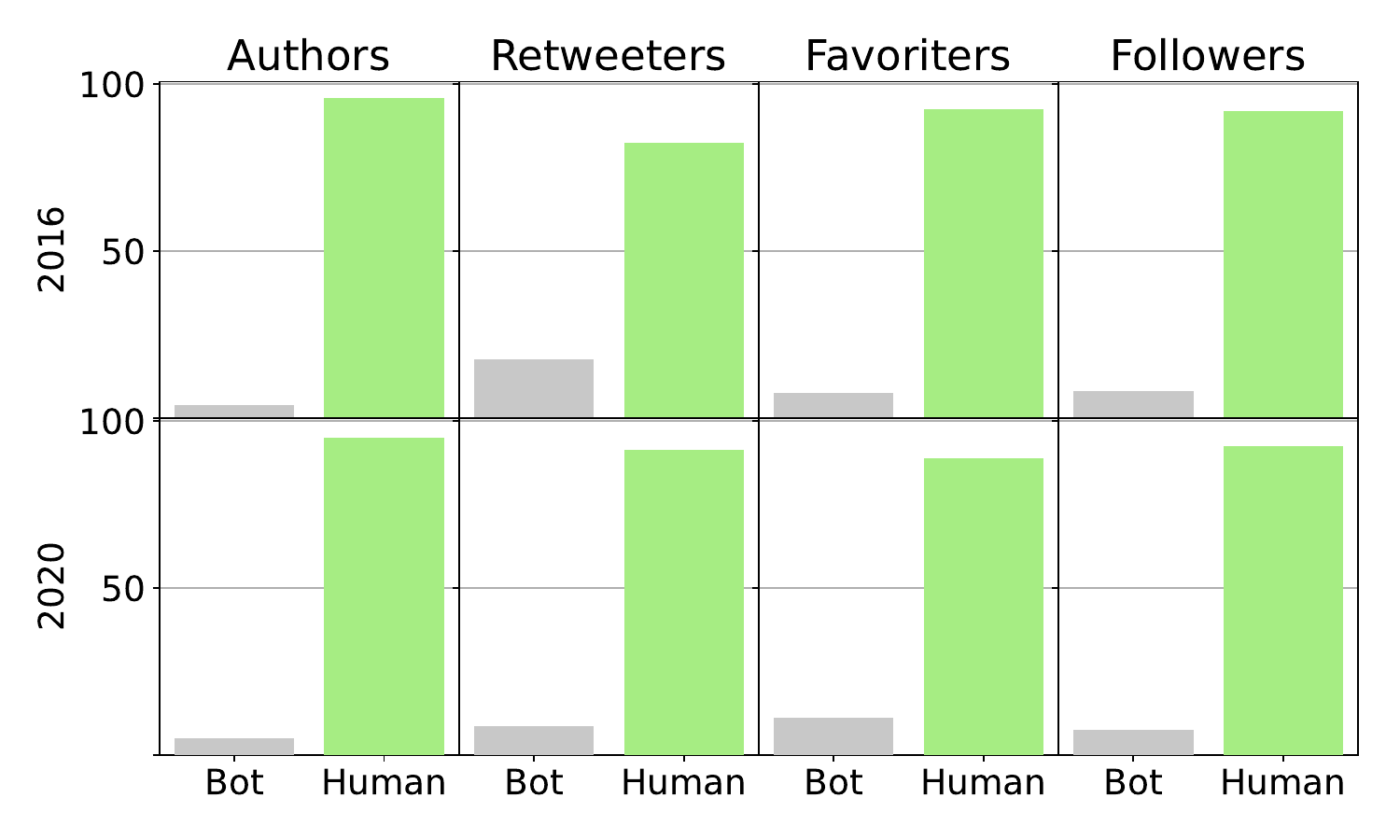}
	\caption{The fraction of accounts that are likely to be bots among poll authors, retweeters, favoriters, and poll author followers. 
 }
	\label{fig:demo-bot}
\end{figure}

We applied the popular bot classifier, \textit{Botometer}, to estimate the fraction of bots among users related to polls. Figure~\ref{fig:demo-bot} plots the fraction of bots separately for poll authors, retweeters, favoriters, and author followers. 
To contextualize these results, we compared the Botometer scores of the authors of presidential-election polls to that of a random sample of Twitter polls posted over the same period and of similar vote distribution. 
Poll authors are not found to score significantly higher on botness than a random sample of non-election poll authors.
In contrast, retweeters of political polls are the highest-scoring user group and indeed are 4 times more likely than authors to be classified as bots (17\% vs. 4\%). This suggests that there may be some degree of astroturfing, i.e., inauthentic user behavior, involved in political campaigning via social polls~\citep{keller2020political}.



\subsection{The relationship between social polling characteristics and poll outcomes}

With the next set of research questions, we aim to uncover potential sources of social poll biases by comparing the characteristics of Twitter polls with their outcomes.
To summarize the results, we regress poll outcomes against the fraction of potential voters of certain political ideology, gender, age, authenticity, and U.S. state~(Table~\ref{tab:regression2020}).

\noindent\textbf{RQ4.1}: How do the results of social polls compare to those of mainstream polls?

Using the results of mainstream social polls and actual vote share as a reference, we describe the poll outcomes. Given Trump's candidacy in both the 2016 and 2020 elections, we present results by referring to the fraction of votes cast in his favor.

\begin{figure}[t!]
	\centering
	\includegraphics[width=0.6\columnwidth]{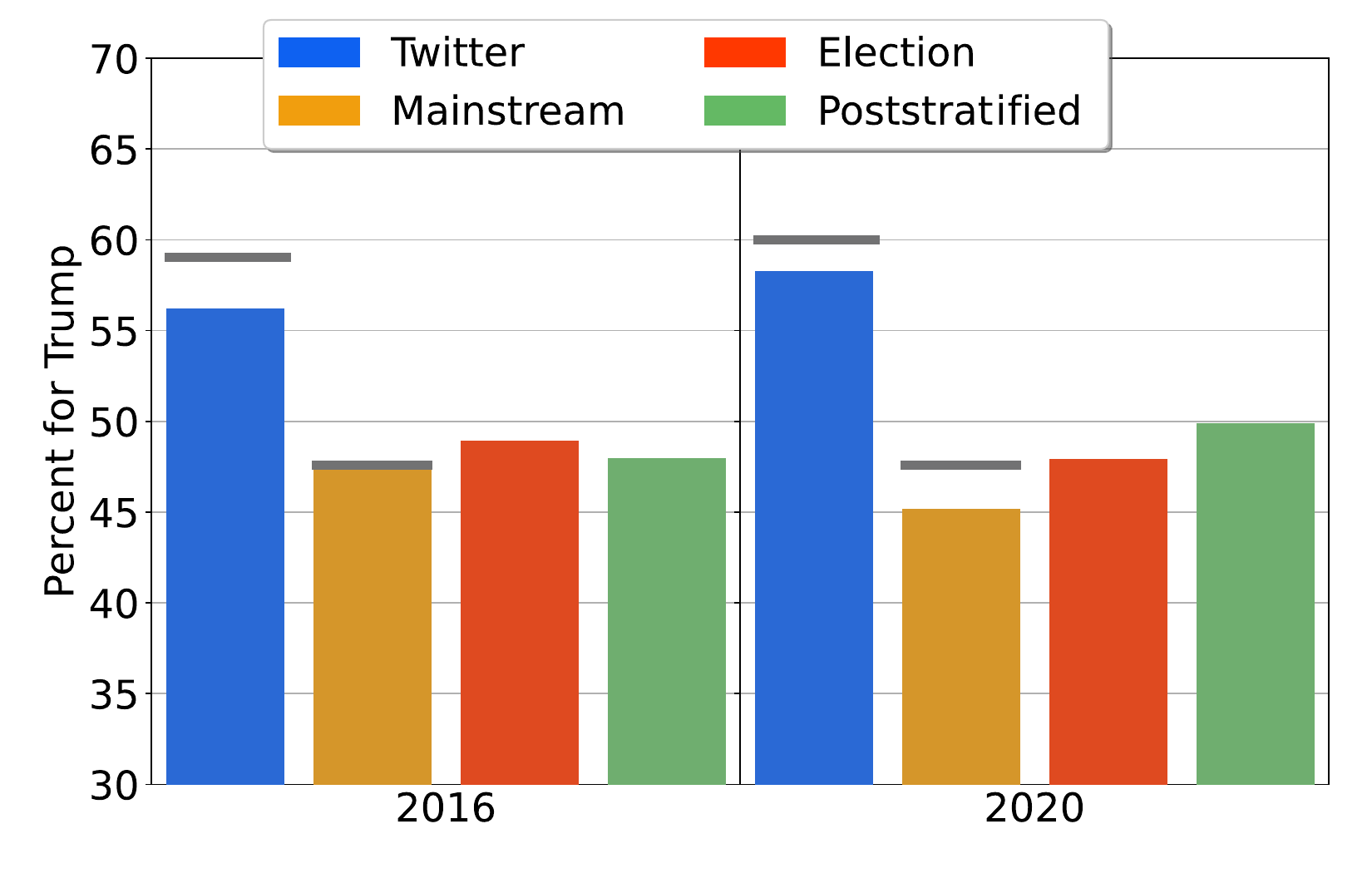}
	\caption{Average outcome of Twitter polls published before the 2016 and 2020 U.S. presidential election days, the average outcomes of mainstream polls, the official election outcomes, and the poststratified Twitter poll outcomes. Gray bars signify the median result.}
    \label{fig:bar-plot}
\end{figure}

Overall, the results of Twitter polls demonstrate that there is a substantial partisan slant, with a consistent leaning toward Trump. 
When analyzing the average Twitter poll results across all polls during the 2016 and 2020 presidential elections (see Figure~\ref{fig:bar-plot}), it becomes evident that these polls tend to overestimate support for Trump compared to actual vote shares on election day. In contrast, mainstream polls tend to underestimate Trump's support. This bias has been reported before, but we note that it is much lower -- and for 2016 disappears -- for normalized traditional polls (compare hatched and unhatched orange and red bars in Figure~\ref{fig:bar-plot}).
Specifically, the median support for Trump in social polls is $17.0 \%p$ higher than that of mainstream polls ($60.0\%$ to $43.0\%$ respectively) in 2020. Mainstream polls, on the other hand, exhibit a bias in favor of the Democratic candidate, with a margin of either 4.74\%p or 5.63\%p when compared to the actual election results. The 2016 election gap between the outcomes of Twitter and mainstream polls is 2\%p larger (59.0\% to 40.27\%), largely the consequence of the well-documented inaccuracies in mainstream polls' election predictions at that time \citep{silver2021death, shiranimehr2018disentangling, gelman2021failure}. 
This contrasting bias pattern suggests that social media polls contain unique information that can be useful for leveling the biases in mainstream polls. 

\begin{table}
\centering
\begin{tabular}{l | l | l }
\toprule
\textbf{Pearson $\rho$}& \textbf{2016} & \textbf{2020} \\
\midrule
\textbf{2-option polls} & -0.12 (p < 0.01) & 0.01 (p = 0.805) \\
\textbf{3-option polls} & -0.34 (p < 0.01) & -0.46 (p < 0.001) \\
\textbf{4-option polls} & -0.59 (p < 0.001) & -0.35 (p < 0.001) \\
\bottomrule
\end{tabular}
\caption{Pearson correlation between the fraction of votes for an option and its position among all potential poll answers, where position 1 is at the top, and position 4 is at the bottom of poll options. The correlation is computed separately for 2-option, 3-option, and 4-option Twitter polls of 2016 and 2020.}
\label{ref:position-outcome}
\end{table}

\noindent\textbf{RQ4.2}: How do poll attributes, such as the order or response options, relate to their outcomes?

The gap between Twitter poll outcomes and actual election results widens in the most popular polls: Trump's lead (measured as a median poll result) accelerates from 60\% for all polls to 69\% for the top 15\% of polls in 2016 and from 60\% to 86\% in 2020, indicating the influence of poll popularity on the bias.
This result is consistent with our other findings that the number of votes correlates with the numbers of retweeters and favoriters who are more likely to be Republican than Democrat (Figure~\ref{fig:demo-political}). 


While the majority of Twitter polls position Trump as the first option among the list of potential poll answers, it is not clear whether this influences poll outcomes. Here, we compute the correlation between the positions and the fractions of votes cast for the candidates placed at the respective positions (Table~\ref{ref:position-outcome}), where position 1 is at the top, and position 4 is at the bottom of the poll options. We find a significant negative correlation between the positions and poll outcomes, suggesting that a candidate who is positioned higher on the list of potential poll answers gets more votes. This correlation, however, is smaller for polls with fewer options, and in the case of head-to-head polls from 2020, the correlation is insignificant.
In fact, the regression of poll outcomes against user attributes and a binary variable encoding whether Trump is above the respective Democratic candidate among poll options reveals no significant impact of the position of poll options on their outcomes (Table~\ref{tab:regression2020}). We explain this result with the prevalence of two-option polls and widespread awareness of the two main candidates in presidential election races. 

\begin{table}[t]
\begin{center}
\begin{tabular}{l|rl|rl}
\multicolumn{1}{c}{} & \multicolumn{2}{c}{\textbf{2016}}& \multicolumn{2}{c}{\textbf{2020}}\\
\toprule
\textbf{Predictor} & \textbf{coef} & \textbf{P$> |$t$|$} & \textbf{coef} & \textbf{P$> |$t$|$}\\
\midrule
{const}                       &       \textbf{0.41}  &    \textbf{***}&  \textbf{0.38}   &    \textbf{***}\\
{$p$(gender=male)}            &   -0.12   &    &   -0.08    & *  \\
{$p$(age$ \ge 40$)}             &   0.10     &  &       \textbf{0.15}  & \textbf{***} \\
{$p$(age$ \in[30,39] $)}       &     0.09  &    & 0.07   &   \\
{$p$(ideology=dem)}           &    -0.10    &     &  -0.15     & * \\
{$p$(ideology=rep)}            &       \textbf{0.33}  &  \textbf{***}&       \textbf{0.38}  & \textbf{***} \\
{$p$(location=blue state)}     &     -0.05   &    &  -0.04    &   \\
{$p$(location=red state)}       &     0.01  &     &     0.02   &   \\
{$p$(bot=yes)}                  &   0.13      &     &    0.05    &   \\
{$p$(first option=Trump)}      &   -0.0003   &      &    -0.03     &   \\
\midrule
\textbf{Dependent variable:} & \multicolumn{2}{l|}{\% for Trump} & \multicolumn{2}{l}{\% for Trump} \\
\textbf{{No. observations:}}      &         \multicolumn{2}{l|}{139}& \multicolumn{2}{l}{641}\\
\textbf{Adj. $\mathbf{R}^2$}:     &         \multicolumn{2}{l|}{0.31}& \multicolumn{2}{l}{0.43}\\
\bottomrule
\end{tabular}
\end{center}
\caption{Parameters of the linear regression model using as dependent variable percent support for Trump in 2016 and 2020 social polls. Predictors encode characteristics of potential voters and poll option ordering information. We indicate statistical significance at levels $p<0.001$ (***), $p<0.01$ (**), and $p<0.05$~(*). Coefficient values in bold font are significant across the two datasets.}
\label{tab:regression2020}
\end{table}

\noindent\textbf{RQ4.3}: How do user attributes relate to the outcomes of social polls?


\begin{figure}[t!]
    \centering
\includegraphics[width=0.99\columnwidth]{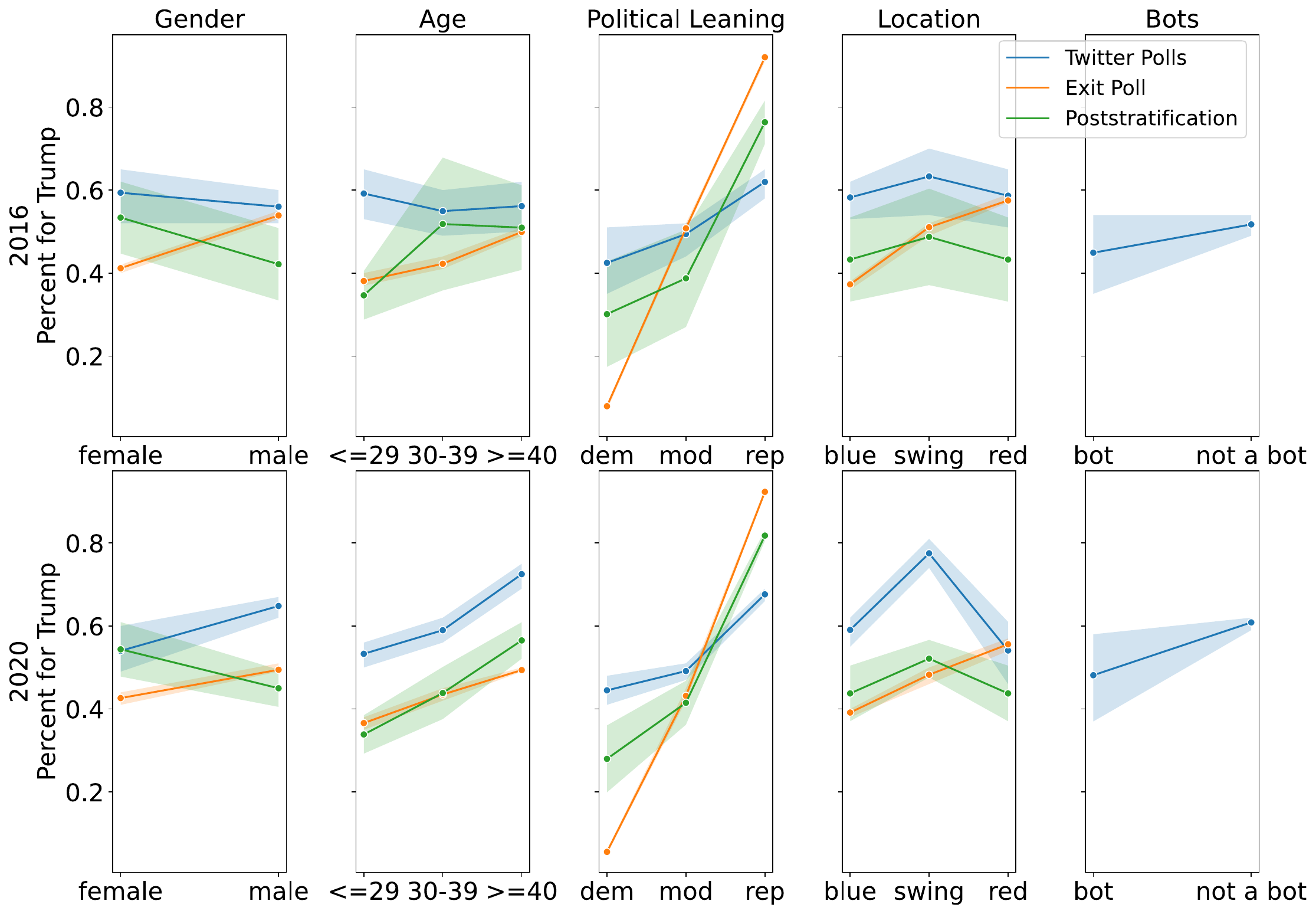}
    \caption{\textbf{}The average fraction of votes for Trump among Twitter polls authored by a user of certain inferred gender, age, and political affiliation (blue) in comparison to the fraction of votes for Trump in respective exit polls (orange) and our poststratified estimates (green). The shaded area marks a 95\% confidence interval.}
    \label{fig:demo_outcome_authors}
\end{figure}

Next, we study whether cross-tabulated results of social polls exhibit similar patterns to the ones of exit polls. For instance, are older or male Twitter users more likely to vote for Trump?
While we do not have direct information about the voters in Twitter polls, it is likely that poll retweeters and favouriters voted in the poll they interacted with. Furthermore, voters in a social poll will likely have similar characteristics to those of the poll's author, due to the effect of social network homophily~\citep{byrne1971attraction, mcpherson2001birds}. For instance, social media users who are connected and interact tend to live close to each other and be of a similar age~\citep{ugander2011anatomy, grabowicz2016road} and share political preferences~\citep{conover2011political, barbera2015tweeting}. 

Remarkably, poll results averaged separately for demographic groups of their authors, that is by age or gender group, reveal the same qualitative demographic cross-tabulation patterns as exit polls (compare blue and orange lines in the left columns of Figure~\ref{fig:demo_outcome_authors}). According to the exit polls, older male voters are more likely to vote for Trump. Similarly, the results of polls authored by older or male users are more biased toward Trump than polls authored by younger or female users.
The bias of social poll outcomes towards Trump in some cases is relatively more pronounced, e.g., among polls authored in 2020 by users 40 years old or older. 

In addition, social poll outcomes show bias toward Trump,
i.e., almost all blue points are above the orange one in Figure~\ref{fig:demo_outcome_authors}.
We explain these results by noting that (i) Republican users have a much higher propensity to author and, even more so, retweet or favorite polls (Figure~\ref{fig:demo-political}), and (ii) there are biases in poll results with respect to political affiliation of social media users (middle column in Figure~\ref{fig:demo_outcome_authors}).
These biases can be addressed by poststratification similar to the one performed for Xbox surveys~\citet{Wang2015}, as we illustrate in the next section.
However, there may be other sources of bias in social media polls, e.g., due to the bias in poll positions, the effect of echo chambers and algorithmic news feeds, or manipulation and campaigning via astroturfing.


We also study how social poll outcomes are related to the locations of users engaged in social polling. To this end, we split U.S. states into blue states, red states, and swing states (Wisconsin, Pennsylvania, New Hampshire, Minnesota, Arizona, Georgia, Virginia, Florida, Michigan, Nevada, Colorado, North Carolina, and Maine).
Interestingly, in 2020 we see a larger support for Trump among users in swing states than in other states (bottom row in Figure~\ref{fig:demo_outcome_authors}), which may correspond to an intensified campaigning by Trump's camp.
However, judging based on the last column of Figure~\ref{fig:demo_outcome_authors}, it is not clear whether polls authored by bot accounts are biased toward Trump more than polls authored by authentic users. 



The regression of social poll outcomes against inferred user attributes and poll option ordering reveals that the two attributes that are the most significantly related to the poll outcomes are the fraction of Republicans and people older than 40 among the potential poll voters (Table~\ref{tab:regression2020}). Older Republican users are likely to vote for Trump.
Democrats are likely to vote for Biden, although the respective coefficient is significant only for 2020, because of the several times smaller sample size for 2016.
The regression models achieve medium adjusted $R^2$ values, $0.43$ and $0.31$ for 2020 and 2016, despite using only a small set of features.
Remarkably, all significant coefficients of the regression models maintain the same sign across the two election years, despite sample size differences, lack of temporal validity (e.g., the political ideology classifier was developed in 2020), and any other potential issues that are not considered in this manuscript. 
We conclude that despite potential manipulation and the impact of algorithmic news feeds and echo chambers, demographic patterns resemble those found in presidential exit polls, suggesting that social polls may contain valuable information about public opinion.

\begin{figure}[t!]
    \centering
    \includegraphics[width=\textwidth]{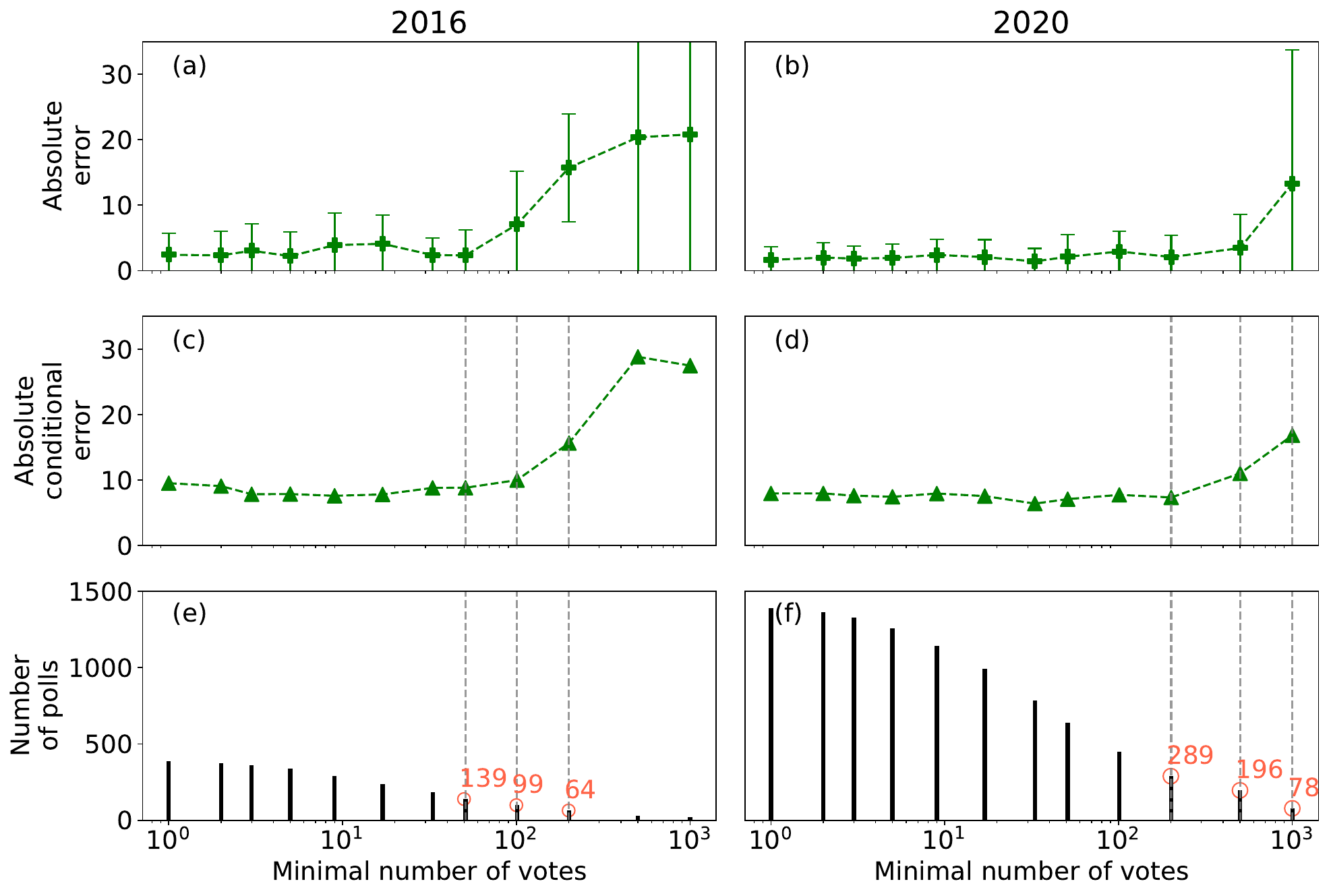}
    \caption{
    The errors of the election outcome estimates grow with the decreasing number of social polls used for poststratification. We filter out the social polls that have less than the number of votes shown on the X-axis.
    (a-b) The absolute error, $|y_e-\hat{y}_e|$, of the election outcome estimate $\hat{y}_e$ based on the poststratified social polls. Error bars mark bootstrapped 95\% confidence intervals.
    (c-d) The average --- across population strata~$g$ --- absolute error of the estimate of support for Trump conditioned on the population stratum $g$.
    (e-f) The number of polls used for poststratification. The vertical dashed lines mark the regions where the errors of estimates begin going~up.
    }
    \label{fig:poststrat_filter_by_votes}
\end{figure}

\subsection{Reducing biases in social poll outcomes to mine public opinion}

We attempt to poststratify the biases in social media poll outcomes by correcting the biases in user attributes of potential voters in social polls. 
The overall poststratification result, estimating the support for the U.S. presidential candidates in the two elections, is shown in Figure~\ref{fig:bar-plot}. 
Strikingly, this approach removes nearly all bias in poll outcomes. 
The poststratified estimate has an error of $0.9\%$ for 2016 and $1.9\%$ for 2020 with respect to the election outcomes. This error is comparable to the error of election forecasts based on traditional polls~\citep{bershidsky2020election, gelman2021failure}.

%

%
Given that the measured error is so low, we explore whether these estimates are stable, rather than being a result of overfitting or randomness.
We note that we obtained this promising result for polls that have at least $M=50$ votes since a lower number of votes translates to a larger variance of the poll outcome. However, it is possible that for other values of $M$ the results of the poststratification are not as good, particularly because this filtering results in less than half of all polls being included in the poststratification (139 and 641 polls for 2016 and 2020, respectively). Will the results be still as good if we use all social polls for poststratification and regression? And what if we use fewer polls?
To answer these questions, we compute the absolute error as a function of the threshold $M$ (Figures~\ref{fig:poststrat_filter_by_votes}a and ~\ref{fig:poststrat_filter_by_votes}b). We find that the error of the poststratified estimate remains low and remarkably stable, between $1\%$ and $4\%$, for all thresholds up to $M=100$ for 2016 and up to $M=500$ for 2020. Also, confidence intervals contain 0 for nearly all of the values of $M$, suggesting that the model is relatively well calibrated, despite its minimalist simplicity. Finally, we note that the number of polls may be more important for accuracy than the threshold $M$, since the errors increase sharply when the number of polls becomes lower than 100 (Figures~\ref{fig:poststrat_filter_by_votes}e and~\ref{fig:poststrat_filter_by_votes}f). 

%
Next, we explore the limits of these promising results. In addition to measuring the accuracy of our estimates at the level of the overall population, we check whether the estimates are accurate also at the level of individual strata.
As ground truth per-stratum election outcomes, we use the exit poll outcomes, $y_e(g)$, conditioned on the stratum $g$. We compute respective conditional estimates based on the social media polls, $\hat{y}_e(g)$ (see Methods), using the same regression models as in the previous paragraph. 
We find that our poststratified conditional results (the green lines in Figure~\ref{fig:demo_outcome_authors}) come significantly closer to the exit poll conditional outcomes (the orange lines) than the raw social poll estimates (the blue lines). Despite this, the two estimates still differ substantially, and about half of the orange points lie outside of the blue confidence intervals. We discuss potential causes in the next paragraph.

We study also how this conditional error depends on the number of polls. To this end, we measure the average (over population strata~$g$) absolute error of the estimate of support for Trump conditioned on the population stratum $g$, that is the average absolute value of the difference between the orange and green points in Figure~\ref{fig:demo_outcome_authors}.\footnote{This is equivalent to $\mathbb{E}_{d\in\mathcal{D}} \mathbb{E}_{g\in \mathcal{G}_d} |y_e(g) - \hat{y}_e(g)|$, where $\mathbb{E}$ corresponds to an expectation.} The average per-stratum error is significantly larger, between $8\%$ and $10\%$ (Figures~\ref{fig:poststrat_filter_by_votes}c and~\ref{fig:poststrat_filter_by_votes}d), than the overall poststratification error that is between $1\%$ and $4\%$. This discrepancy may have a number of causes. 
First, the regression model may be not specified correctly, e.g., it misses interaction terms between the predictors and does not explicitly account for the heteroscedasticity of the data (the variance of poll outcomes depends on the number of votes). Second, the definitions of the dimensions characterizing voters may differ between our user attribute classifiers and exit poll codebooks, e.g., political ideology may be defined in slightly different ways.
Third, the exit poll estimates may be biased estimates of the ground truth conditional support for the candidates.
Either way, the absolute conditional error increases with the decreasing number of polls in the same way as the overall poststratification error, so the two measures are clearly related, as one may have expected.

We conclude that to make accurate estimates of public opinion, it is not enough to have a dozen polls with a large number of votes, let alone one poll, as was the case for the Literary Digest's 1936 survey. However, if we have a hundred biased polls with their demographic crosstabs, accurate estimates turn out to be possible at least at the level of the overall voting population.


\section{Discussion}

Our examination of Twitter polls offers insights at multiple levels. First, this study provides a comprehensive description of how Twitter polls were used during the 2016 and 2020 U.S. presidential elections. Second, our findings suggest various sources of biases present in Twitter polls. Third, this study paves the way for future research in social polling.

This study suggests that Twitter polls could be an effective tool for boosting political engagement. Platforms like Twitter have the potential to cultivate a sense of community and belonging in the online political sphere, thereby stimulating greater political participation and voter mobilization~\citep{bond201261millionperson, rojas2009mobilizers}.
Our study primarily focuses on identifying and understanding the factors that could introduce biases in social media polls, such as the presentation order of poll options, the spread of polls through homogeneous networks, and the influence of algorithmic news feeds. 
We notice a pattern of increased political engagement among Trump supporters, indicated by their active participation in sharing and interacting with Twitter polls. This pattern of engagement mirrors the pattern observed in 2020 where there was a 20\% gap in \textit{strong} support for Trump over Biden \citep{pew2020elect}. The findings are also consistent with what previous research suggests \cite{Wells2016}: Trump voters are more inclined to engage with social media content related to his candidacy.

We highlight the risks associated with undisclosed biases in social poll outcomes, which can act as a source of polarizing and misleading content~\citep{wu2019misinformation}. As seen in other studies~\citep{vosoughi2018spread, juul2021comparing, zafar2016message, brady2019ideological}, misleading and polarizing Twitter polls garner more attention and spread more widely than true and balanced information. 
Such misinformation often resonates more with individuals who have strong partisan identities, exacerbating the spread of polarizing content~\citep{babaei2019analyzing, pennycook2021psychology, guess2021cracking, allcott2017social}. Likewise, polarizing misinformation, such as biased polls, can gain more support among people with strong partisan identities.
This phenomenon is reflected in the asymmetric distribution of partisanship among users engaging with misinformation~\citep{nikolov2021right, gonzalezbailon2023asymmetric}, a pattern also evident in those interacting with social polls.

Furthermore, social polls offer a common, structured format to compare expressions of political preference across a variety of platforms with diverse user bases, scopes, and values, and thus enable an unprecedentedly nuanced characterization of public opinion and political engagement.
While differences exist between social and mainstream polls, their parallels suggest a potential for future studies to apply statistical methods to correct biases in social polling. For instance, \citet{Wang2015} demonstrated that with the appropriate statistical adjustments of demographic and political variables, non-representative polls from Xbox users can yield accurate election forecasts. Our findings underline the potential of social media in capturing public opinion once biases in data are corrected.  
While we achieve very low poststratification errors at the overall population level, future user attribute inference and poststratification approaches could establish social polls as a vital data source for estimating public opinion, either independently or in conjunction with mainstream polls.
Future studies can use much larger datasets and more complex models, e.g., accounting for heteroscedasticity and interaction terms between predictors.
Overall, social media polls 
may become a valuable source of information about a wide breadth of political preferences. They offer a promising alternative to measuring public opinion from social media over sentiment analysis, which has had limited success so far~\citep{oconnor2010tweets, jungherr2015, schober2016} and faces multiple challenges due to its reliance on unstructured information in the form of text~\citep{klasnja2018measuring, dong2021review, diaz2016online}.

Overall, this study opens up multiple avenues for future research. These include exploring the motivations behind publicly sharing political opinions in social polls, comparing audience perceptions of social polls versus institutional surveys, and their impact on offline political mobilization. 
While our study is focused on U.S. presidential races, future research could expand to other political elections, both domestic and international.
Twitter (re-branded as X in 2023) is, at the time of writing, a primary avenue where the public and elite advance the political discourse in the U.S., and thus was a natural setting for studying social polls about presidential candidates. Yet, social polling as a form of political activity is a global phenomenon that is commonplace on several online platforms beyond Twitter. For instance, polls can be created on Facebook Pages and Groups. Correspondingly, the opportunities for research are far greater than what has been covered by the present study. Understanding how social polling varies across platforms, and how it is used across a wide range of topics, is necessary to comprehend social polling as a broader practice in an age of interconnected media.

\section{Limitations}

This study performs inference of several unobserved attributes of Twitter users, such as their age, gender, and political leaning. We recognize that, although we used the most accurate inference methods available and evaluated them on our data, such inference is still subject to error. Therefore, the results of the inference are used only in aggregate and should be interpreted as coarse-grained comparisons between user cohorts.

We note also that for a poll to be in our dataset, it must contain both "vote" and the candidates' names in its text, which is not exhaustive.
Twitter API changes cut researchers' access to Twitter data in 2023 when this study was conducted. In light of this newly imposed restriction limiting Twitter's transparency, we deemed it important to publish the results of this study despite the potential incompleteness of the data, to keep the public informed about Twitter polls that could have impacted the voting behavior of U.S. citizens and could have reinforced the beliefs of a subset of the U.S. citizens in voting fraud. 
Subsequent work can identify usable and precise methods to cut through the vast noise in social platforms for a more complete analysis of relevant polls. 


\section{Broader Perspective, Ethics, and Competing Interests}

This research obtained approval from the pertinent Institutional Review Board. 
This work concerns itself with the political opinions of Twitter users. As such, it is important to discuss the ethical aspects of the fielding and dissemination of this research. 

First, we believe that social media platforms should be transparent about user activity in relation to political issues such as elections. We note that biased social polls can be misleading and can contribute to voter fraud belief. As such, it is important that social media platforms inform users about biases in political polls

Second, the computational model used for inference affords a specific and limited operationalization of gender as a binary construct. This operationalization is admittedly limited, but instrumental in comparing the results of this research to existing research hypotheses and data from mainstream polls that also adopt it. The authors recognize the need for further studies that endorse more complex conceptualizations of user characteristics, and especially that perform at-scale surveying of unobserved user characteristics, although such efforts exceed the scope of this first-comer analysis.


\section{Acknowledgments}
We thank Filippo Menczer and Kaicheng Yang for providing us with the Botometer scores for millions of Twitter users. We also thank Pablo Barberá for sharing the estimated political leanings of Twitter users.

\bibliography{biblio, social_public_opinion}


\end{document}